\documentclass[12pt, draftclsnofoot, onecolumn]{IEEEtran}

\def\ignore#1{}

\usepackage{amsmath,epsfig,epsf,psfrag,amssymb,amsfonts,latexsym,amsmath,color}

\usepackage{caption}
\usepackage{epstopdf}
\DeclareGraphicsExtensions{.eps}

\usepackage{dsfont}
\usepackage{verbatim}
\usepackage{cite,enumerate}
\usepackage{srcltx,cases}
\usepackage[mathscr]{eucal}

\usepackage{relsize}

\newtheorem{lemma}{Lemma}
\newtheorem{proposition}{Proposition}
\newtheorem{corollary}{Corollary}
\newtheorem{remark}{\textit{Remark}}

\newtheorem{example}{Example}

\newcommand{\var}{\textnormal{Var}}
\newcommand{\cov}{\textnormal{Cov}}

\newcommand{\beq}{\begin{equation}}
\newcommand{\eeq}{\end{equation}}

\begin{document}

\title{ A 3-D Spatial Model for In-building Wireless Networks with Correlated Shadowing }

\author{  \IEEEauthorblockN{Junse Lee, Xinchen Zhang, and Fran\c{c}ois Baccelli} 
	}

\date{}
\maketitle

\begin{abstract}

Consider orthogonal planes in the 3-D space representing floors and walls
in a large building.
These planes divide the space into rooms where a wireless infrastructure is
deployed. This paper is focused on the analysis of the
correlated shadowing field created by this wireless infrastructure 
through the set of walls and floors.
When the locations of the planes and of the wireless nodes
are governed by Poisson processes,
we obtain a simple stochastic model which captures the non-uniform
nature of node deployment and room sizes.
This model, which we propose to call
the \emph{Poisson building}, 
captures the complex in-building shadowing correlations,
is scalable in the number of dimensions and is tractable
for network performance analysis.
It allows an exact mathematical characterization of the interference
distribution in both infinite and finite buildings,
which further leads to closed-form expressions for
the coverage probabilities in in-building cellular networks
and the success probability of in-building underlay D2D transmissions.
\end{abstract}

\IEEEpeerreviewmaketitle
\section{Introduction}\label{sec:intro}

In-building wireless networking is predicted to be one of the
fastest growing markets of the wireless industry.
Since traffic increase is expected to come from indoor networks, mobile operators are investigating in-building network deployment in recent and upcoming years\cite{ciscowhite2015}. The potential of the in-building wireless market 
largely comes from the complement it offers to conventional outdoor
network deployments, and from
the exponential growth of mobile traffic demand.
On the other hand, the spatial modeling of in-building wireless
networks largely remains an uncharted area despite
the great progress in the planar (2-D) modeling of wireless
networks over the past decade.
This work presents a first attempt toward getting
a tractable comprehensive 3-D spatial model in this context.

\subsection{Main Contributions: the Poisson Grid Model}

Compared with classical planar models, the main technical challenge
for obtaining a tractable 3-D spatial model for in-building
wireless networks lies in the proper handling of the shadowing 
correlation created by the static physical objects which shape the
way wireless signal propagates and attenuates over the Euclidean space.
As most of the planar models are designed to study outdoor networks
at the scale of a city, the shadowing correlation is typically ignored 
and path loss is simply modeled through independent log-normal shadowing coefficients\cite{ilow1998analytic,blaszczyszyn2015wireless} or distance-dependent
function combined with independent fading/shadowing random variables\cite{bai2014analysis,BaiHeath2015}.
In contrast, in-building networks are typically much denser and 
heavily shadowed by physical objects (floors and walls).
The scale of these objects is comparable and often much larger than inter-node distances,
resulting in highly correlated shadowing in space.

This work presents the \emph{Poisson grid} model,
which explicitly handles the shadowing correlation,
which we regard as one of the main challenges of in-building network modeling.
The \emph{Poisson grid} is also dimension-scalable in that it can be constructed and analyzed for 2-D, 3-D  and
even higher dimensional networks in a consistent fashion (Sect. \ref{subsec:Poisson_Grid}, \ref{subsec:Transmitters_Poisson_building}).
The prominent application of the Poisson grid is its \emph{3-D incarnation}, also referred as the \emph{Poisson building},
which is particularly useful to study the performance of 3-D wireless networks in large buildings.
This model is compatible with the empirically
supported lognormal shadowing model in that the marginal shadowing
component converges to lognormal distribution 
as the link distance grows (Sect. \ref{subsec:Path_loss}).

We demonstrate the tractability of the Poisson building
model by explicitly deriving the interference distribution 
and its spatial correlation (Sect. \ref{sec:typical_room}, \ref{sec:typical_user}).
This in turn leads to analytical characterizations of
the success probability (SINR distributions) 
of D2D underlay networks and the coverage probability of in-building cellular networks (Sect. \ref{sec:Success and Coverage Probability}).
Finally, we briefly touch on a couple of important variants, namely
the finite Poisson building and
the semi-infinite Poisson building (Sect. \ref{sec:finite_size}).
The latter allows one to analyze the interference in a window office,
which is a boundary office in a large semi-infinite building.
The analysis of these variants further reveals fundamental
differences between 3-D and 2-D correlated shadowing analysis.

\subsection{Related Works}

\subsubsection{3-D Network Models}

The 3-D Poisson building model is not the only 3-D model
for wireless networks; nor the simplest one.
One obvious alternative is to generalize the usual 2-D model directly
by distributing nodes as a Poisson point process in the 3-D space
and applying a distance-based path
loss function\cite{GuptaKumar2001,GuptaZhangAndrews2015}. 
This will be referred to as the \emph{free-space 3-D model}.
While such a model is analytically convenient, it may appear to
be oversimplified in some contexts.
Distance-based path loss models are usually derived using free space
propagation assumptions (Friis' equation) and a simplified
ground reflection model (e.g., the 2-ray model 
or the Hata model\cite{GoldsmithBook,ZhangAndrews2015}). 
Therefore, applying this model to the 3-D in-building context amounts
to ignoring the major path loss contributor, namely (spatially correlated) blockage.
In contrast, the Poisson building model is built in order to represent blockage effects and to provide a compact mathematical
model for in-building networks with variable size rooms.

The free-space 3-D and the Poisson building are compared 
in Sect. \ref{subsec:two_model_comparison}, where important metrics pertaining to the distribution
of interference created by the very same collection of wireless
nodes are shown to lead to arbitrarily large discrepancies.

\subsubsection{Ray-tracing}

Accurate in-building network analysis can be achieved by the 3-D ray-tracing\cite{mckown1991ray,Valenzuela1993}.
As a site-specific approach, 3-D ray-tracing requires sophisticated software packages and an exact building geometry.
In contrast, the stochastic geometric modeling approach of this paper is based on analyzing a random structure of obstacles with a small number of key parameters.
It thus works without a complete description of the propagation environment and is
more flexible in obtaining general design guidelines for 3-D in-building networks.
In fact, in the long term, this approach might provide a
theoretically justified and rather simple alternative to 3-D
ray-tracing software platforms which are often difficult
to build and use.

\subsubsection{Correlated Shadowing}
As in \cite{GoldsmithBook}, shadowing is highly correlated.
Yet, few generative or tractable models have been proposed to cope with this.
The first model of correlation was proposed by Gudmundson
\cite{gudmundson1991correlation} to model the lognormal shadowing
variable between a fixed base station (BS) and a moving user
by an autoregressive process with an exponentially decaying autocorrelation.
As a result, the spatial dependence of shadowing is formulated by joint Gaussian distributions.
The multi-base station (BS)\cite{graziosi2002general} and multi-hop network
\cite{agrawal2009correlated} cases were also considered based on similar ideas.
This approach also forms the basis of the model suggested by the
3GPP\cite{access2010further} and 802.11 standardization groups\cite{erceg2004tgn}.

These models have two main weaknesses. First, it is hard to give a clear physical interpretation to the joint Gaussian distribution used to
model spatially correlated shadowing. Second, the models have limited
tractability for large dense wireless networks. 
Complex simulation platforms need to be set up to implement these models.
In contrast, the model presented in this paper has a clear (blockage-penetration) physical interpretation
and is tractable in deriving important performance metrics in closed forms.

\subsubsection{Stochastic Geometry and Shadowing Models}
Stochastic geometric models have become popular for the analysis
of spectral efficiency in wireless networks for both D2D and
cellular networks\cite{net:ElSawy13tut}. Independent shadowing 
fields can be incorporated into the basic models
\cite{ilow1998analytic,blaszczyszyn2015wireless,bai2014analysis}. However, correlated
shadowing fields have not yet been combined with stochastic geometric models.
Recently, we started developing a log-normal compatible model for
analyzing the correlation structure in 2-D urban networks
\cite{BaccelliZhang2015,ZhangBaccelliHeath2015,Lee1604:Shadowing}.
This paper uses the same methodology and extends it to high dimensional indoor networks (with 3-D being the main application).
The analytical performance characterization presented this paper shows that
the 3-D case enjoys a comparable tractability and provides a unified framework for networks with arbitrary dimension.

\section{System Model}\label{sec:system model}
\subsection{The Poisson Grid}\label{subsec:Poisson_Grid}

\begin{figure}
	\begin{center}
		\epsfxsize=3.8in {\epsfbox{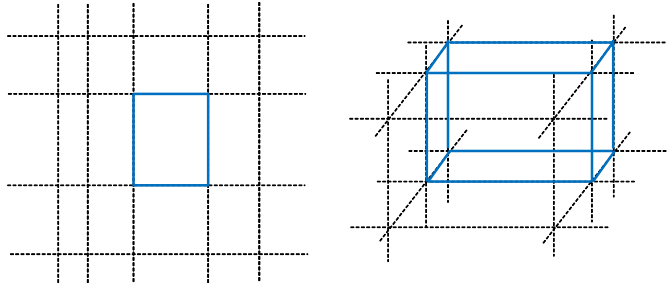}}
		\caption{\small One realization of Poisson grids for 2-D and 3-D. Typical room $(0,0,\ldots,0)$ is highlighted with a solid box. }
		\label{fig:network model}
	\end{center}
\end{figure}

The Poisson grid is constructed on the $n$-dimensional\footnote{Below, $n$ will be 2 or 3, but since there is
no cost handling the general case, we keep $n$ general in 
the model and most of the derivation.} Euclidean
space $\mathbb{R}^n,~n\in\mathbb{N}\cap[2,\infty)$.
It consists of a collection of (hyper-)planes perpendicular to
the axes of the Euclidean space. This is a generalization of the (2-D)
Manhattan Poisson Line Process (MPLP)\cite{net:Daley-VereJones07}.
We consider $n$ Cartesian axes and name them $v_1,v_2,\ldots,v_n$.
We build independent homogeneous Poisson Point Processes (PPP)
along the $v_1,v_2,\ldots,v_n$-axis, with intensities
$\mu_1,\mu_2,\ldots,\mu_n$, respectively. At each point of these
processes, an infinite hyperplane grows perpendicular to the 
axis on which the point is located. We denote this random structure
by $\Psi = \bigcup_{i=1}^n\Psi_i$,
where $\Psi_i$ is the collection of hyperplanes grown from the points
on $v_i$. This divides the space into infinitely many rectangular boxes or rooms.
Fig. \ref{fig:network model} gives an example of the 2-D (MPLP)
and the 3-D (Poisson building) cases.

\subsection{Transmitters on Room Corners and Ceiling Lines}\label{subsec:Transmitters_Poisson_building}

To reflect realistic network deployments, we assume that all the transmitters (infrastructure nodes, also referred as BSs) are located on some of the one dimensional facets of $\Psi$, as in \cite{BaccelliZhang2015,Lee1604:Shadowing}. This is inspired by the fact that most real-life wireless infrastructure nodes (small cell BSs or WiFi access points) are mounted along ceiling lines or placed at corners of rooms. Since each intersection line segment meets $2^{n-1}$ rooms, we build $2^{n-1}$ independent transmitter processes along each line segment and assign them to the adjacent rooms. On the lines parallel to $v_i$ ($i\in[n]$)\footnote{We use [n] to denote the set $[1,n]\cap\mathbb{N}$.}, the transmitters are distributed as a homogeneous PPP with intensity $\lambda_i$.\footnote{Assuming each of the adjacent $2^{n-1}$ processes having the same density $\lambda_i$ is only for convenience and can be easily generalized. In fact, as will become obvious later in the paper, all of the results will stay the same if we apply different densities $\lambda_{i,1},\lambda_{i,2},\dots,\lambda_{i,2^{n-1}}$ to these processes but keep $2^{n-1}\lambda_i = \sum_{j=1}^{2^{n-1}} \lambda_{i,j}$. }
The resulting point process (transmitters) is denoted by $\Phi$, which is a stationary Cox point process in $\mathbb{R}^n$.  The mean number of BSs per unit volume\footnote{This will be used in Sect.\ref{subsec:two_model_comparison} to compare SIR distribution of our Cox point process model in the Poisson grid and the previous PPP models in \emph{free-space}. (\ref{eq:cox_avg_lambda}) is the ratio of the mean number of BSs in one room to the mean size of a room. } is
\begin{equation}\label{eq:cox_avg_lambda}
\textstyle \lambda_{avg}=\left(\sum_{i=1}^n\frac{2^{n-1}\lambda_i}{\mu_i}\right)\left(\prod_{j=1}^n\mu_j\right)\mbox{.}
\end{equation}

\subsection{Path Loss Model}\label{subsec:Path_loss}
\subsubsection{Blockage-Based Path Loss Model}
We consider a \emph{blockage-based} path loss model, where the received signal power at $y$ from the transmitter at $x$ ($x,y\in\mathbb{R}^n$) is \begin{align}\label{eq:path_loss_model}
P_{x\rightarrow y}= P_{tx}h\prod_{i=1}^n K_i^{N_i},
\end{align}
where $P_{tx}$ is the received power of a same room communication (\emph{i.e.,} both transceivers are in the same room) without fading, $h$ is the i.i.d. channel fading coefficient between $x$ and $y$, $K_i\in[0,1),i\in[n]$ is the penetration loss of the hyperplanes perpendicular to the $v_i$ axis, and $N_i,i\in[n]$ is the number of hyperplanes grown from the point process on the $v_i$ axis between $x$ and $y$. To be precise, $N_i = |\overline{xy}\cup\Psi_i|$, where $\overline{xy}$ is the open line segment connecting $x$ and $y$ and $|\cdot|$ denotes the cardinality of a set. Without loss of generality, we assume $P_{tx} = 1$, which does not affect the SINR distribution after proper rescaling of the thermal noise power.

One possible concern on this model is the absence of distance-based path loss term. This is justified by the fact that blockage dominates distance-based loss in indoor environments, which aligns with intuition and is corroborated by ray-tracing studies\cite{marano2005ray}.

\subsubsection{Compatibility with Log-normal Shadowing}
For an arbitrary link $x\rightarrow y$ with given Euclidean length $\|x-y\|=d$, and angle (w.r.t. $v_i, i\in[n]$) $\vartheta_i$, $N_i$ is Poisson distributed with mean $\mu_i d \cos(\vartheta_i)$. Thus, the path loss can be rewritten as $\exp\left(-\sum_{i=1}^{n}N_i\log(\frac{1}{K_i})\right)$ where $\log(\frac{1}{K_i})>0$. As $\mu_i\rightarrow \infty$ or as $d\rightarrow \infty$, $N_i$ can be well approximated by a normal random variable. In other words, combining a blockage-based path loss model and the Poisson grid indoor geometry creates a marginal shadowing distribution which is lognormal, and thus connects the model with the data supported lognormal shadowing.

\subsection{Coverage and Success Probability}
This paper considers two communication scenarios. The first is a \emph{cellular downlink} scenario, where we focus on deriving the \emph{coverage probability} $\mathbb{P}[{\sf SINR}_c>\theta]$, where
\begin{align}
{\sf SINR}_c \triangleq \frac{P_{d\rightarrow r}}{\sum_{t\in\Phi\setminus\{d\}}P_{t\rightarrow r}+\sigma^2}.\nonumber
\end{align}
Here, $d$ is the serving BS, $r$ the receiver, and $\sigma^2$ the thermal noise power. This is the probability that a chosen user observes an SINR higher than a threshold $\theta$. As a function of $\theta$, $\mathbb{P}[{\sf SINR}_c>\theta]$ can be interpreted as the complementary cumulative distribution function (CCDF) of SINR.

We also consider a \emph{D2D underlay} scenario, where a mobile user attempts to connect to another user using the (shared) cellular spectrum. We analyze the \emph{success probability}, $\mathbb{P}[{\sf SINR}_s>\theta]$, where
\begin{align}
{\sf SINR}_s \triangleq \frac{P_{link}}{\sum_{t\in\Phi}P_{t\rightarrow r}+\sigma^2},\nonumber
\end{align}
with $P_{link}$ being the received power of the target D2D link.
\section{Interference in the Typical Room}\label{sec:typical_room}

Define the total interference as the sum of the received power from all transmitters. When the channel coefficients are $h\equiv 1$ (i.e., without fading), the interference is the same at any point of a given room according to our model. In this section, we focus on the interference in the typical room (formally defined below), and give the moments and the distribution of the total interference. 

Precisely, we consider the intersection points of $n$-orthogonal planes. We denote this stationary point process by $\xi$ and consider the Palm version of $\xi$. Under its Palm version, $\xi$ has a point at the origin of $\mathbb{R}^n$. Denote by (0, 0,\ldots, 0) the room which contains this point and is in the positive orthant.
We refer to this room as \emph{the typical room}, and label the other rooms by their relative position with respect to the typical room. Intuitively, the typical room is a uniformly randomly chosen room in the Poisson grid.

In the case without fading, we denote the interference observed in room $(i_1,i_2,\ldots,i_n)$ by $I_{(i_1,i_2,\ldots,i_n)}\triangleq \sum_{x\in\Phi}P_{x\rightarrow r}$ where $r$ is any point in room $(i_1,i_2,\ldots,i_n)$.

\subsection{Interference Moments}
\proposition[Mean Interference]\label{prop:mean_room_lines} In the absence of fading ($h\equiv 1$), the mean interference observed in the typical room is
\begin{align} \textstyle\mathbb{E}[I_{(0,0,\ldots,0)}]=2^{n-1}\left(\sum_{j=1}^n\frac{\lambda_j}{\mu_j}\right)\left(\prod_{i=1}^n\frac{1+K_i}{1-K_i}\right).\nonumber
\end{align}
\begin{IEEEproof}
Let $N_{(i_1,i_2,\ldots,i_n)}$ be the number of the BSs in room $(i_1,i_2,\ldots,i_n)$. Denote the side lengths of this room by $d_{1_{i_1}}, d_{2_{i_2}},\ldots,d_{n_{i_n}}$ where $d_{j_{i_j}},(j\in [n])$ are independent exponential random variables with mean $\frac{1}{\mu_j}$, denoting the length of the side parallel to the $v_j$-axis. For a given structure $\Psi$, 
\begin{align}
\textstyle\mathbb{E}[N_{(i_1,i_2,\ldots,i_n)}]=\mathbb{E}[\mathbb{E}[N_{(i_1,i_2,\ldots,i_n)}|\Psi]]=\mathbb{E}[2^{n-1}\sum_{j=1}^n\lambda_jd_{j_{i_j}}]=2^{n-1}\sum_{j=1}^n\frac{\lambda_j}{\mu_j}.\nonumber
\end{align}
Since the attenuation from room $(i_1,i_2,\ldots,i_n)$ to the typical room is $\prod_{j=1}^nK_j^{|i_j|}$\mbox{,} 
\begin{align*}
\textstyle\mathbb{E}[I_{(0,0,\ldots,0)}]=\sum_{(i_t)_{t=1}^n\in\mathbb{Z}^n}\mathbb{E}[N_{(i_1,i_2,\ldots,i_n)}]\prod_{j=1}^nK_j^{|i_j|}=2^{n-1}\left(\sum_{j=1}^n\frac{\lambda_j}{\mu_j}\right)\left(\prod_{i=1}^n\frac{1+K_i}{1-K_i}\right)\mbox{.}
\end{align*}
The last step comes from the fact that $\sum_{i\in\mathbb{Z}}K^{|i|}=\frac{1+K}{1-K}$.
\end{IEEEproof}

\begin{example}
When $n=2$, the mean interference observed in the typical room reduces to { $2\left(\prod_{i=1}^2\frac{1+K_i}{1-K_i}\right)\left(\sum_{i=1}^2\frac{\lambda_i}{\mu_i}\right)$}. When $n=3$, it becomes  $4\left(\prod_{i=1}^3\frac{1+K_i}{1-K_i}\right)\left(\sum_{i=1}^3\frac{\lambda_i}{\mu_i}\right)$. In the 3-D case, when there are no BS along the $v_3$ axis, (i.e. $\lambda_3=0$), the ratio of the interference in the 3-D typical room to that of the 2-D typical room is $2\left(\frac{1+K_3}{1-K_3}\right)$. The factor $2$ comes from the fact that there are twice more BSs in any of the $v_1,v_2$ directions in the 3-D model (\emph{e.g.,} those on the ceiling lines, and those on the floor lines); $\frac{1+K_3}{1-K_3}$ reflects the interference leaked from other floors.
\end{example}

\begin{figure*}[t]
	\begin{minipage}{0.49\linewidth}
		\begin{center}
			\epsfxsize=2.8in {\epsfbox{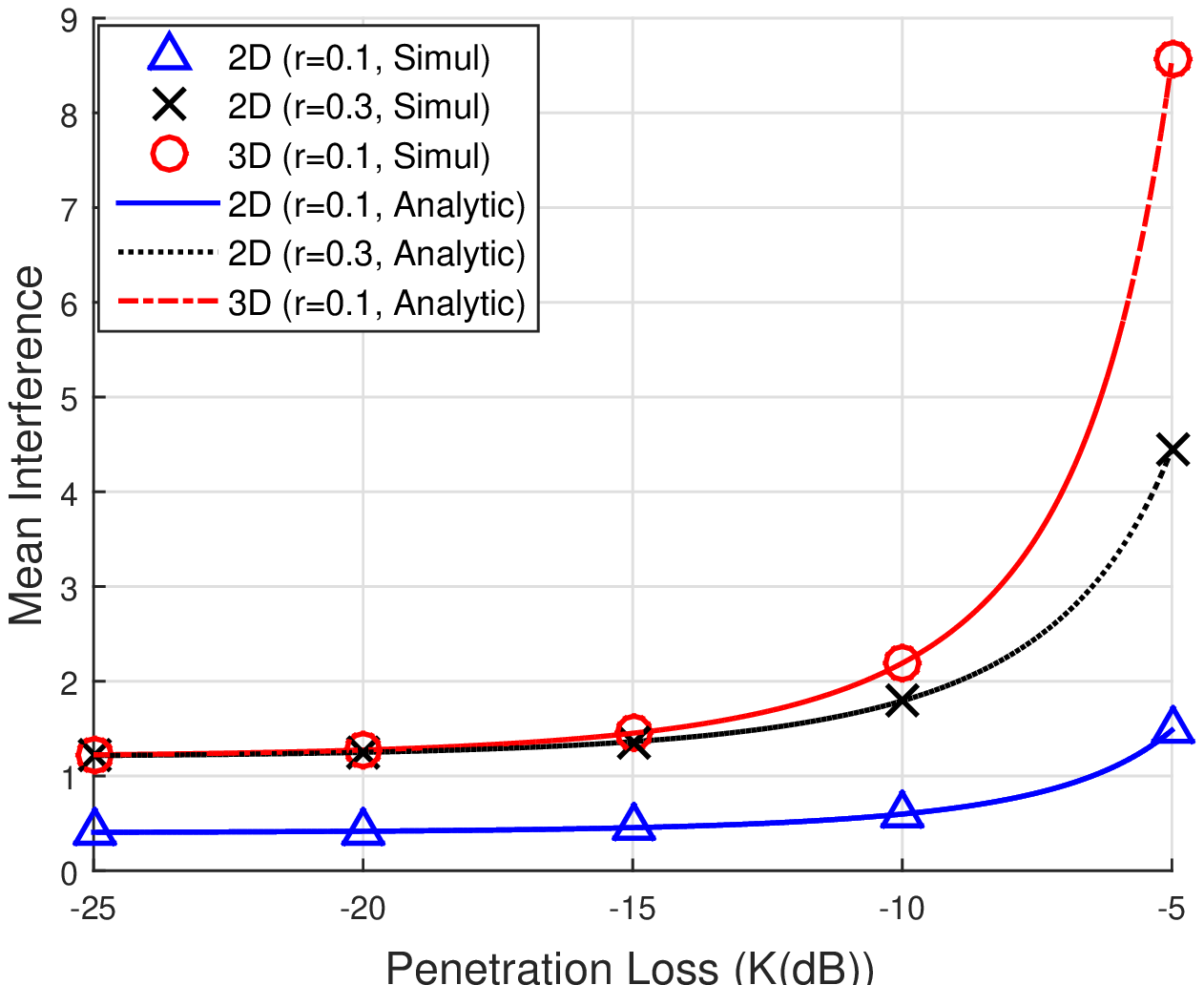}}
			\caption{\small Mean interference where $r_i=\frac{\lambda_i}{\mu_i}=r$, and $K_i=K$ for $i=1,2$ in the 2-D case and $i=1,2,3$ in the 3-D case }
			\label{fig:mean_interference}
		\end{center}
	\end{minipage}
	\hspace{0.01\linewidth}
	\begin{minipage}{0.49\linewidth}
		\begin{center}
			\epsfxsize=2.8in {\epsfbox{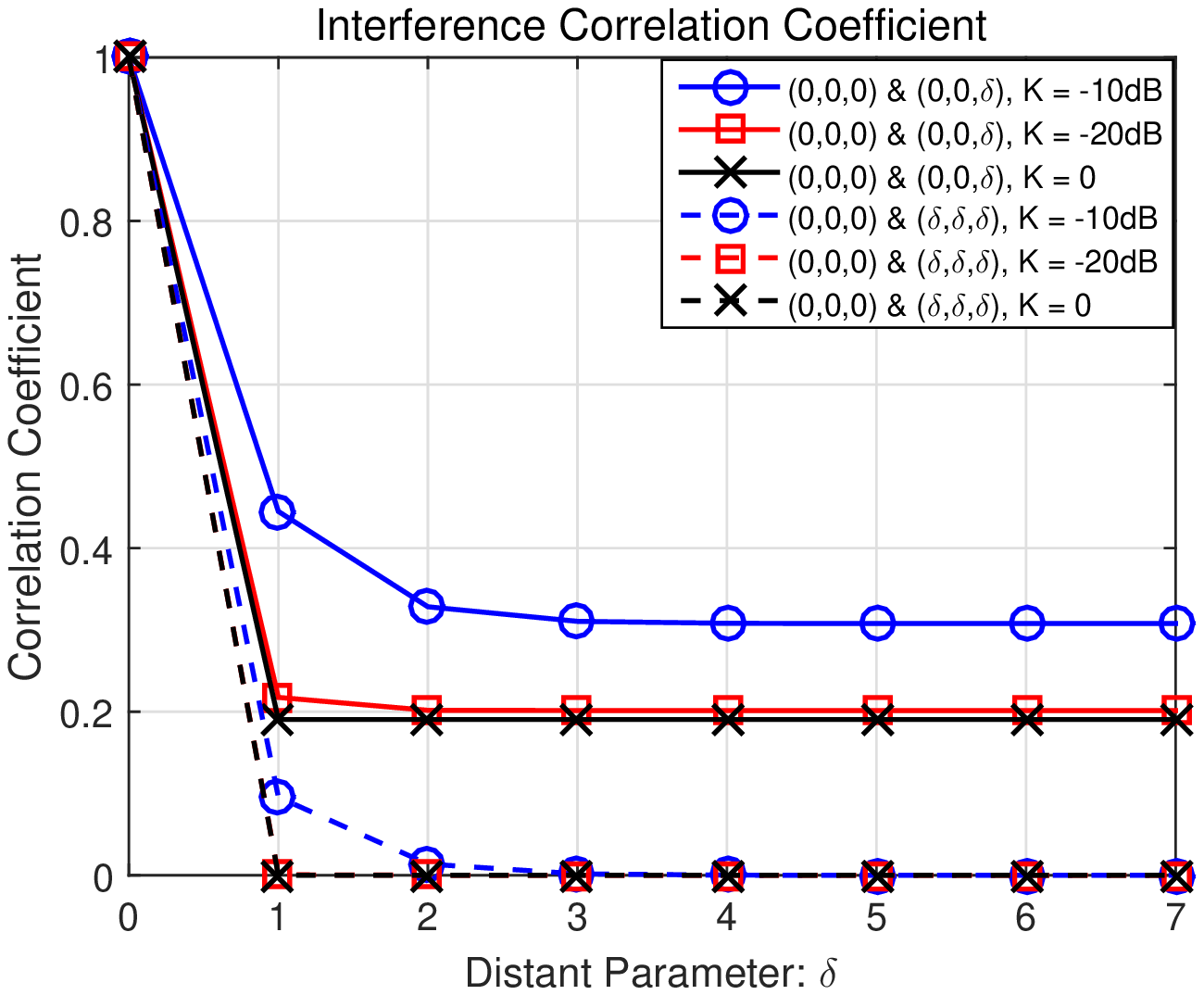}}
			\caption{\small Interference Correlation Coefficient between the typical room and room $(0,0,\delta)$ ($=\rho_{(0,0,\delta)}$) and the typical room and room $(\delta,\delta,\delta)$ ($=\rho_{(\delta,\delta,\delta)}$) in the 3-D case where $K_i=K$ and $r_i=\frac{\lambda_i}{\mu_i}=0.1$ for $i=1,2,3$}
			\label{fig:corr_coeff}
		\end{center}
	\end{minipage}
\end{figure*}

Fig. \ref{fig:mean_interference} illustrates the mean interference observed by the typical room in the 2-D and the 3-D cases. We assume all penetration losses are the same (i.e., $K_i=K,~ \forall i$) and the ratios of the transmitter density to the wall density are identical (i.e. $r_i=\frac{\lambda_i}{\mu_i}=r,~ \forall i$). Since there are more edges in a higher dimensional room, the mean interference of 3-D is larger than that of 2-D, under the same $K$ and $r$. Also, as $K$ decreases, the mean interference in 3-D decreases faster compared to that of 2-D. As $K$ goes to 0, the hyperplanes shield the interference from other rooms perfectly. So, when $K\rightarrow 0$, the mean interference converges to the mean number of the transmitters in the typical room of the Poisson grid.

\proposition[Interference Joint Moment]\label{prop:moment_room_lines} In the absence of fading (i.e., $h\equiv 1$), the joint moment of the interference between the typical room and room $(i_1,i_2,\ldots,i_n)$ is {
\begin{align*}
	\textstyle\mathbb{E}[I_{(0,0,\ldots,0)}I_{(i_1,i_2,\ldots,i_n)}]&\textstyle= 2^{n-1}\left(\sum_{j=1}^n\frac{\lambda_j}{\mu_j}\right)\left(\prod_{l=1}^nb_l(i_l)\right)\\
	&\textstyle+2^{2n-2}\left(\prod_{l=1}^na_l\right)\left(\left(\sum_{j=1}^n\frac{\lambda_j}{\mu_j}\right)^2 +\left(\sum_{j=1}^n\frac{\lambda_j^2b_j(i_j)}{\mu_j^2a_j}\right)\right),
\end{align*}}where 
\begin{align*}
\textstyle a_i=\left(\frac{1+K_i}{1-K_i}\right)^2, b_i(x)=K_i^{|x|}\left(|x|+\frac{1+K_i^2}{1-K_i^2}\right),
\end{align*}
for $i\in[n]$.\begin{IEEEproof}
	See Appendix \ref{appendix:moment_room_lines}.	
	\end{IEEEproof}	\corollary[Interference Variance]\label{cor:interference_variance} By Propositions \ref{prop:mean_room_lines} and \ref{prop:moment_room_lines} with $(i_1,i_2,\ldots,i_n)=(0,0,\ldots,0)$, the variance of the interference observed in the typical room is{	
	\begin{align*}
	\textstyle \var[I_{(0,0,\ldots,0)}]=	2^{n-1}\left(\sum_{j=1}^n\frac{\lambda_j}{\mu_j}\right)\left(\prod_{l=1}^nb_l(0)\right) +2^{2n-2}\left(\sum_{j=1}^n\frac{\lambda_j^2b_j(0)}{\mu_j^2a_j}\right)\left(\prod_{l=1}^na_l\right).\end{align*}}\remark Due to the stationarity of the Poisson grid, the correlation coefficient between the typical room and room $(i_1,i_2,\ldots,i_n)$ is $\rho_{(i_1,i_2,\ldots,i_n)}=\cov[I_{(0,0,\ldots,0)},I_{(i_1,i_2,\ldots,i_n)}]/\var[I_{(0,0,\ldots,0)}]$.

Fig. \ref{fig:corr_coeff} shows the interference correlation coefficient in the 3-D case ($\rho_{(0,0,\delta)}$ and $\rho_{(\delta,\delta,\delta)}$) where $\delta\in\mathbb{N}\cup\{0\}$. As expected, when the penetration loss $K$ goes from -10dB to 0 (-$\infty$dB), $\rho_{(0,0,\delta)}$ and $\rho_{(\delta,\delta,\delta)}$ decrease. Furthermore, 1) $\rho_{(0,0,\delta)}$ does not go to zero when $K=0$ (\emph{i.e.,} no interference leakage between rooms), and 2) $\rho_{(0,0,\delta)}$ does not go to zero even if $\delta$ goes to infinity. Both observations can be explained by the correlation of the room sizes along the corresponding axis directions. Intuitively, a large room is more likely next to a large room due to the shared building frame. On the other hand, $\rho_{(\delta,\delta,\delta)}$ goes to zero if $K$ goes to zero and $\delta$ goes to infinity as the typical room and room $(\delta,\delta,\delta)$ do not share side(s). This intricate behavior of interference correlations highlights the impact of room size correlation in a typical in-building environment. This impact is well manifested in Fig.~\ref{fig:corr_coeff}, but is impossible to capture using conventional (free-space) models.

\remark[Scale-invariance] By Propositions \ref{prop:mean_room_lines}, \ref{prop:moment_room_lines} and Corollary \ref{cor:interference_variance}, the interference moments of two Poisson grids (\emph{i.e.,} with different $\lambda_i,~\mu_i$) are identical if the ratios of the transmitter density to the wall density $r_i=\frac{\lambda_i}{\mu_i}, i\in[n]$ as well as the penetration losses $K_i, i\in[n]$ are identical.
\subsection{Interference Distribution}
\proposition[Interference Distribution without Fading]\label{prop:laplace_dist_no_fading_lines} Without fading (i.e., $h\equiv 1$), the Laplace transform of the interference observed in the typical room is 
\begin{align*}
\textstyle\mathcal{L}_{I_{(0,0,\ldots,0)}}(s)=\prod_{k=1}^{n}f\left(s,\frac{\lambda_k}{\mu_k},\left(K_{(i+k)\%n}\right)_{i=0}^{n-1}\right),
\end{align*}
where $m\%n$ denotes $m$ modulo\footnote{In this paper, we consider that the range of modular operation by an integer $N$ is $1$ to $N$.} $n$, and {
\begin{align*} 
\textstyle f\left(s,x,\left(K_{i}\right)_{i=1}^{n}\right)&\textstyle=f(s,x,K_1,K_2,\ldots,K_n)\\\textstyle&\textstyle=\prod_{i_1\in\mathbb{Z}}\left({1+2^{n-1}x\sum_{(i_p)_{p=2}^n\in\mathbb{Z}^{n-1}}(1-\exp(-s\prod_{q=1}^n K_q^{|i_q|}))}\right)^{-1}.
\end{align*}}\begin{IEEEproof} The Laplace transform of the interference given $\Psi$ is
	\begin{align*} \textstyle\mathcal{L}_{I_{(0,0,\ldots,0)}|\Psi}(s)=\prod_{k=1}^{n}f_c\left(s,{\lambda_k},\left(K_{(i+k)\%n}\right)_{i=0}^{n-1}\right),
	\end{align*}	where
	$f_c\left(s,x,\left(K_{i}\right)_{i=0}^{n-1}\right)=\prod_{i_1\in\mathbb{Z}}\left({2^{n-1}x\sum_{(i_p)_{p=2}^n\in\mathbb{Z}^{n-1}}(1-\exp(-s\prod_{q=1}^n K_q^{|i_q|}))}\right).$
 Since $d_{ij}$ are i.i.d. exponential random variables, we obtain the Laplace transform by deconditioning w.r.t. the Poisson grid.
	\end{IEEEproof}
	
Let $\tilde{I}_{(i_1,i_2,\ldots,i_n)}$ be the interference in room $(i_1,i_2,\ldots,i_n)$ where the channel is subject to Rayleigh fading.
	
\proposition[Fading]\label{prop:laplace_dist_fading_lines} Under Rayleigh fading ($h\sim \exp(1)$), the Laplace transform of the interference observed in the typical room is {
\begin{align*}
\textstyle\mathcal{L}_{\tilde{I}_{(0,0,\ldots,0)}}=\prod_{k=1}^{n}\tilde{f}\left(s,\frac{\lambda_k}{\mu_k},\left(K_{(i+k)\%n}\right)_{i=0}^{n-1}\right),
\end{align*}}where {
\begin{align*} 
\textstyle\tilde{f}\left(s,x,\left(K_{i}\right)_{i=1}^{n}\right)&\textstyle= \tilde{f}(s,x,K_1,K_2,\ldots,K_n)\\\textstyle&\textstyle=\prod_{i_1\in\mathbb{Z}}\left({1+2^{n-1}x\sum_{(i_p)_{p=2}^n\in\mathbb{Z}^{n-1}}(1-\frac{1}{1+s\prod_{q=1}^n K_q^{|i_q|}})}\right)^{-1}.
\end{align*}}\begin{IEEEproof}
	The proof is analogous to that of Proposition \ref{prop:laplace_dist_no_fading_lines}, except for the fact that the interference from room $(i_1,i_2,\ldots,i_n)$ is the sum of i.i.d. exponential random variables with mean $\prod_{m=1}^nK_m^{|i_m|}$.
	\end{IEEEproof}

We also provide the joint interference distribution at two rooms. Characterizing the joint distribution is important for analyzing the Quality of Service (QoS) of users when they travel across rooms and is non-trivial under the previous stochastic geometric models.

  \proposition[Joint Laplace Transform]\label{prop:joint_laplace_transform} Under Rayleigh fading (i.e., $h\sim \textnormal{Exp}(1)$), the joint Laplace transform of the interference in the typical room and in room $(l_1,l_2,\ldots,l_n)$ is {
\begin{align*}
  \textstyle\mathcal{L}_{\tilde{I}_{(0,0,\ldots,0)}\tilde{I}_{(l_1,l_2,\ldots,l_n)}}(s_1,s_2)=\prod_{k=1}^{n}\tilde{f}_{\sf j}\left(s_1,s_2,\frac{\lambda_k}{\mu_k},\left(K_{(i+k)\%n}\right)_{i=0}^{n-1},\left(l_i\right)_{i=0}^{n-1}\right),
\end{align*}}where {
\begin{align*} 
\textstyle&\textstyle\tilde{f}_{\sf j}\left(s_1,s_2,x,\left(K_{i}\right)_{i=1}^{n},\left(l_i\right)_{i=1}^{n}\right)=\tilde{f}_{\sf j}(s_1,s_2,x,K_1,K_2,\ldots,K_n,l_1,l_2,\ldots,l_n)\\\textstyle&\textstyle=\prod_{i_1\in\mathbb{Z}}\left({1+2^{n-1}x\sum_{(i_p)_{p=2}^n\in\mathbb{Z}^{n-1}}(1-\frac{1}{1+s_1\prod_{q=1}^n K_q^{|i_q|}}\frac{1}{1+s_2\prod_{q=1}^n K_q^{|i_q-l_q|}})}\right)^{-1}\mbox{,}
\end{align*}
and the subscript ${\sf j}$ stresses the \emph{joint} distribution.}

The proof follows the line of thought in \cite{ganti2009spatial} and is omitted.
\section{Interference at a Typical User}\label{sec:typical_user}

At the beginning of Section \ref{sec:typical_room}, we defined the point process $\xi$ and discussed its Palm distribution. This section is focused on the stationary distribution case or equivalently takes the perspective of the typical user. The typical user is located at the origin of the $n$-dimensional Euclidean space. Without fading (i.e., $h\equiv 1$), the interference at the typical user is denoted by $I_o\triangleq\sum_{x\in\Phi}P_{x\rightarrow o} $, where $o$ is the origin. and denoted by $\tilde{I}_o$ under Rayleigh fading.

\begin{figure}
	\begin{center}
		\epsfxsize=2.6in {\epsfbox{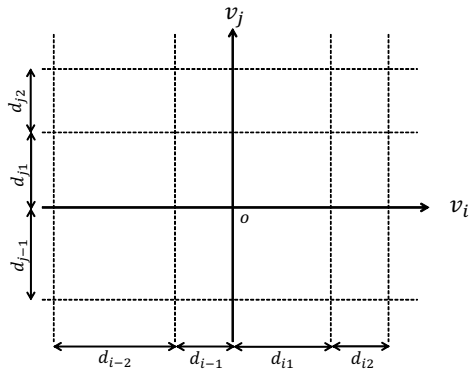}}
		\caption{\small Labeling system for Section \ref{sec:typical_user}  }
		\label{fig:typical_user}
	\end{center}
\end{figure}

As indicated in Fig. \ref{fig:typical_user}, we use a different labeling system. The main difference between this labeling system and the one in Section~\ref{sec:typical_room} is that the room containing $o$ is divided into $2^n$ pseudo rooms. By construction, each of the pseudo rooms has identically, exponentially distributed sides.

\subsection{Interference Distribution}

\proposition[No fading]\label{prop:laplace_dist_no_fading_lines_user} Without fading, i.e., $h\equiv 1$, the Laplace transform of the interference $I_o$ observed by a typical user is {
\begin{align*}
\textstyle\mathcal{L}_{I_o}(s)=\prod_{k=1}^{n}g\left(s,\frac{\lambda_k}{\mu_k},\left(K_{(i+k)\%n}\right)_{i=0}^{n-1}\right),
\end{align*}}where {
\begin{align*}  
\textstyle g\left(s,x,\left(K_{i}\right)_{i=1}^{n}\right)&\textstyle=g(s,x,K_1,K_2,\ldots,K_n)\\
&\textstyle=\prod_{i_1\in\mathbb{N}}\left({1+2^{n-1}x\sum_{(i_p)_{p=2}^n\in\mathbb{Z}^{n-1}}(1-\exp(-s\frac{\prod_{q=1}^n K_q^{|i_q|}}{K_{1}}))}\right)^{-2}.
\end{align*}}\begin{IEEEproof}
	The proof is analogous to that of Proposition \ref{prop:laplace_dist_no_fading_lines}. The main difference is that the edges of the typical room have lengths distributed like the sum of two exponential random variables.
	\end{IEEEproof}	By differentiating the formula of Proposition \ref{prop:laplace_dist_no_fading_lines_user}, we obtain the following result.
	
	\proposition\label{prop:mean_user_line} In the absence of fading, the mean interference observed at the typical user is 
	\begin{align*}
	\textstyle\mathbb{E}[I_{o}]=2^{n-1}\left(\prod_{i=1}^n\frac{1+K_i}{1-K_i}\right)\left(\sum_{j=1}^n\frac{\lambda_j}{\mu_j}\frac{2}{1+K_j}\right).
	\end{align*}
	
\remark[$n$-D Feller's Paradox\cite{feller2008introduction}] By comparing Propositions \ref{prop:mean_room_lines} and \ref{prop:mean_user_line}, the amount of interference observed by the typical user is larger than the interference in the typical room. This result comes from the fact that the size of the typical room is smaller than the room containing the typical user, which makes the user ``see'' a larger number of strong (near) interferers. More formally, the zero-cell (cell which contains the origin) is chosen with a size bias with respect to the typical cell under Palm distribution, and this favors larger cells, which have in turn more chance to cover a fixed point.

\proposition[Interference Distribution with Fading]\label{prop:laplace_dist_fading_lines_user} Under Rayleigh fading, the Laplace transform of the interference $\tilde{I}_o$ at the typical user is {
\begin{align*}
\textstyle\mathcal{L}_{\tilde{I}_o}(s)=\prod_{k=1}^{n}\tilde{g}\left(s,\frac{\lambda_k}{\mu_k},\left(K_{(i+k)\%n}\right)_{i=0}^{n-1}\right),
\end{align*}}where {
\begin{align*} 
\textstyle\tilde{g}\left(s,x,\left(K_{i}\right)_{i=1}^{n}\right)&\textstyle=\tilde{g}(s,x,K_1,K_2,\ldots,K_n)\\
\textstyle&\textstyle=\prod_{i_1\in\mathbb{N}}\left({1+2^{n-1}x\sum_{(i_p)_{p=2}^n\in\mathbb{Z}^{n-1}}(1-\frac{K_{1}}{K_{1}+s{\prod_{q=1}^n K_q^{|i_q|}}})}\right)^{-2}.
\end{align*}}\begin{IEEEproof}
	The proof is analogous to that of Proposition \ref{prop:laplace_dist_fading_lines} and is omitted.
\end{IEEEproof}

\subsection{Comparison of Correlated and Uncorrelated Shadowing}\label{subsec:corr_uncorr}

In classical stochastic geometric models, the shadowing coefficients of different links are modeled using i.i.d. log-normal random variables\cite{ilow1998analytic,blaszczyszyn2015wireless} or depend only on the lengths of each link\cite{bai2014analysis}. In these models, the shadowing correlation is typically ignored. In this subsection, we compare the statistical differences between our correlated model and the distance-based uncorrelated shadowing model.
We will focus on the 3-D case\footnote{It is possible to generalize this to the $n$-dimensional case.}, and denote the interference observed by the typical user under correlated and uncorrelated shadowing by
$I_{o,cor}$ and $I_{o,unc}$, respectively.
\subsubsection{Poisson Grid with \emph{Correlated} and \emph{Uncorrelated} Shadowing}

For a fair comparison, we analyze the uncorrelated case with the same Cox node distribution as in the Poisson grid model.
That is the transmitters are also distributed on the lines of a Poisson grid made of planes parallel to the axes.
In the uncorrelated model, the penetration losses of the transmitters are independently sampled from the marginal distribution of the number of walls that block their link.

\begin{figure}
	\begin{center}
		\epsfxsize=3.5in {\epsfbox{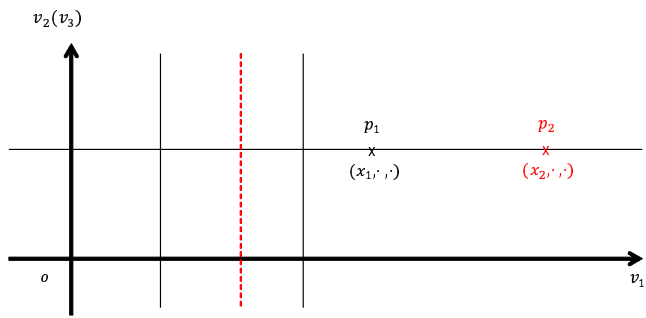}}
		\caption{\small Uncorrelated Shadowing model. The number of $v_1$-orthogonal walls from the origin to $p_1$ is $Poiss(\mu_1x_1)$, and to $p_2$ is $Poiss(\mu_1x_2)$. With a positive probability, $Poiss(\mu_1x_1)$ can be larger than $Poiss(\mu_1x_2)$.}
		\label{fig:uncorrelated_model}
	\end{center}
\end{figure}

Let us first focus on the nodes on the lines parallel to $v_1$ and denote these nodes by $\Phi_{v_1}$. For each transmitter $x\in\Phi_{v_1}$ with $v_1$-coordinate $x_{v_1}$, the number of planes orthogonal to $v_1$ between this transmitter and the typical user is a Poisson random variable with mean $\mu_1|x_{v_1}|$. To analyze the difference between the correlated and the uncorrelated models, we pick two transmitters $p_1$ and $p_2$ with $v_1$-coordinates $x_1$ and $x_2$ respectively and such that $0<x_1<x_2$.  In the correlated model, there are always fewer $v_1$-orthogonal walls between $x_1$ and the typical user than between $x_2$ and the typical user, a property which is not guaranteed under the uncorrelated model (See Fig. \ref{fig:uncorrelated_model}.) As we will see, the difference can result in non-trivial discrepancies in interference statistics.

\subsubsection{Mean Interference}
From Proposition \ref{prop:mean_user_line}, we have
\corollary\label{cor:corr_mean} Without fading, in the 3-D case, the mean interference at the typical user under correlated shadowing is
			\begin{align*}
			\textstyle\mathbb{E}[I_{o,cor}]=4\left(\prod_{i=1}^3\frac{1+K_i}{1-K_i}\right)\left(\sum_{j=1}^3\frac{\lambda_j}{\mu_j}\frac{2}{1+K_j}\right)\mbox{.}
			\end{align*} 
\proposition\label{prop:uncorr_mean} Under the 3-D uncorrelated shadowing model without fading, the mean interference observed by the typical user is
	\begin{align}\label{eq:uncorr_mean}
	\textstyle\mathbb{E}[I_{o,unc}]=\mathbb{E}[I_{o,cor}]\mbox{.}
	\end{align}
	\begin{IEEEproof}
		For $x\in\Phi_{v_1}$ with $v_1$ coordinate $x_{v_1}$, the expectation of the power attenuation by $v_1$-orthogonal walls is 
\begin{align*}
\sum_{n=0}^{\infty}K_1^n\frac{(\mu_1|x_{v_1}|)^n}{n!}e^{-\mu_1|x_{v_1}|}=e^{-\mu_1|x_{v_1}|(1-K_1)}\mbox{.}
\end{align*}
Since $\Phi$ is the union of independent PPPs, we compute the interference moments from one PPP and aggregate the contributions of all PPPs.
By Campbell's formula\cite{baccelli2009stochastic}, in the uncorrelated case, the mean interference from the transmitters on one $v_1$-parallel line such that that there is no $v_2$ and $v_3$ orthogonal wall between this line and the origin is{
\begin{eqnarray*}
	\textstyle\mathbb{E}[I_{\Phi_{v_{11}}}]=\lambda_1\int_{\mathbb{R}}e^{-\mu_1|x|(1-	K_1)}dx=\lambda_1\frac{2}{\mu_1(1-K_1)}\mbox{,}
\end{eqnarray*}}when we denote these transmitters by $\Phi_{v_{11}}$ and the interference from $\Phi_{v_{11}}$ by $I_{\Phi_{v_{11}}}$. If we aggregate all transmitters on the $v_1$-parallel lines, the mean interference is
\begin{align*}
\textstyle \mathbb{E}[I_{\Phi_{v_1}}]
&\textstyle=\mathbb{E}[I_{\Phi_{v_{11}}}] \times 4\left(\sum_{i\in\mathbb{Z}}K_2^{|i|}\right)\left(\sum_{j\in\mathbb{Z}}K_3^{|j|}\right)\\&\textstyle=\lambda_1\frac{2}{\mu_1(1-K_1)}4\prod_{i=2}^3\left(\frac{1+K_i}{1-K_i}\right)\\
\textstyle&\textstyle=4\left(\prod_{i=1}^3\frac{1+K_i}{1-K_i}\right)\left(\frac{\lambda_1}{\mu_1}\frac{2}{1+K_1}\right)\mbox{.}
\end{align*}We obtain (\ref{eq:uncorr_mean}) by using the same line of thought, for all transmitters (including transmitters on the lines parallel to the $v_2$-axis and the $v_3$-axis).
\end{IEEEproof}

\subsubsection{Variance}
From the formula of Proposition \ref{prop:laplace_dist_no_fading_lines_user}, we get:
\corollary For the correlated shadowing case, in the absence of fading, the variance of the interference observed by the typical user in the 3-D case is
\begin{eqnarray*}
	\textstyle \var[I_{o,cor}]=\left(\prod_{i=1}^3\frac{1+K_i}{1-K_i}\right)^2\left(\sum_{j=1}^3\frac{32(1-K_j)}{(1+K_j)^3}\frac{\lambda_j^2}{\mu_j^2}\right)+\left(\prod_{i=1}^3\frac{1+K_i^2}{1-K_i^2}\right)\left(\sum_{j=1}^3\frac{8}{1+K_j^2}\frac{\lambda_j}{\mu_j}\right)\mbox{.}
\end{eqnarray*}

\proposition For the 3-D uncorrelated shadowing case, the variance of interference is
\begin{align*}
\textstyle \var[I_{o,unc}]=4\left(\prod_{i=1}^3\frac{1+K_i^2}{1-K_i^2}\right)\left(\sum_{j=1}^3\frac{\lambda_j}{\mu_j}\frac{1+K_j}{1+K_j^2}\right)\mbox{.}
\end{align*}\begin{IEEEproof}
The expectation of the square of the interference from $\Phi_{v_{11}}$ is 
\begin{eqnarray*}
	\textstyle\mathbb{E}[I_{\Phi_{v_{11}}}^2]
	&&\textstyle=\mathbb{E}\left[\left(\sum_{X\in\Phi_{v_{11}}}e^{-|X|\mu_1(1-K_1)}\right)^2\right]\\&&\textstyle=\mathbb{E}\left[\left(\sum_{X\in\Phi_{v_{11}}}e^{-|X|\mu_1(1-K_1)}\right)\left(\sum_{Z\in\Phi_{v_{11}}}e^{-|Z|\mu_1(1-K_1)}\right)\right]\\
	&&\textstyle=\mathbb{E}\left[\left(\sum_{X\in\Phi_{v_{11}}}e^{-2|X|\mu_1(1-K_1)}\right)\right]+\mathbb{E}\left[\sum_{X,Z\in\Phi_{v_{11}}}^{X\neq Z}e^{-(|X|+|Z|)\mu_1(1-K_1)}\right]\\
	&&\textstyle=\lambda_1\int_{\mathbb{R}}e^{-2|x|\mu_1(1-K_1)}dx+\lambda_1^2\int_{\mathbb{R}}\int_{\mathbb{R}}e^{-(|x|+|z|)\mu_1(1-K_1)}dx dz\\
	&&\textstyle=\lambda_1\int_{\mathbb{R}}e^{-2|x|\mu_1(1-K_1)}dx+\left(\lambda_1\int_{\mathbb{R}}e^{-|x|\mu_1(1-K_1)}\right)^2\mbox{,}
\end{eqnarray*}
where we used the fact that the second factorial moment measure of a PPP is the Lebesgue measure \cite{baccelli2009stochastic}.
So, the variance of interference from $\Phi_{v_{11}}$ is
\begin{equation*}
\textstyle \var[I_{\Phi_{v_{11}}}]=\lambda_1\int_{\mathbb{R}}e^{-2|x|\mu_1(1-K_1)}dx=\frac{\lambda_1}{\mu_1(1-K_1)}\mbox{.}
\end{equation*}
Since the PPPs on different lines are independent, we obtain that the variance of the interference from transmitters on the $v_1$-parallel lines is
\begin{align*}
\textstyle \var[I_{\Phi_{v_{1}}}] &=\textstyle \var[I_{\Phi_{v_{11}}}]\times 4\left(\sum_{i\in\mathbb{Z}}K_2^{2|i|}\right) \left(\sum_{j\in\mathbb{Z}}K_3^{2|j|}\right)\\&\textstyle=\frac{\lambda_1}{\mu_1(1-K_1)}4\left(\frac{1+K_2^2}{1-K_2^2}\right)\left(\frac{1+K_3^2}{1-K_3^2}\right)\mbox{.}
\end{align*}
\end{IEEEproof}
 
\remark In general, the variances of the correlated and uncorrelated shadowing cases are different.
If we assume $\lambda_2,\lambda_3=0$, the variance ratio between the correlated and uncorrelated cases can be simplified into
\begin{align}\label{eq:compa_var_corr_uncorr}
\textstyle\frac{\var[I_{o,cor}]}{\var[I_{o,unc}]}&\textstyle=\frac{2}{1+K_1}\left(1+\frac{4\lambda_1}{\mu_1}\frac{(1+K_2)^3}{(1-K_2)(1+K_2^2)}\frac{(1+K_3)^3}{(1-K_3)(1+K_3^2)}\right).
\end{align}
Equation \eqref{eq:compa_var_corr_uncorr} shows that the tail of $I_{o,cor}$ is heavier than that of $I_{o,unc}$, which aligns with the observation in \cite[Corollary 3]{ZhangBaccelliHeath2015}. As $\frac{\lambda_1}{\mu_1}$ decreases and $K_1$ approaches 1,\footnote{$K_1$ can be arbitrarily close to but not equal to 1, as the derivation used the fact that $\sum_{j\in\mathbb{Z}}K_1^{|j|}=\frac{1+K_1}{1-K_1}$.} the variance ratio goes to 1. Intuitively this is explained by the fact that in Fig. \ref{fig:uncorrelated_model}, if there is no penetration loss through $v_1$-orthogonal walls, and the probability that the path loss between the origin and $p_1$ is larger than that between the origin and $p_2$ becomes $0$. For a special case, when $K_2,K_3=0$, (i.e., all $v_2$, $v_3$-orthogonal walls totally shield the signal stemming from the next rooms), the ratio becomes $\frac{2}{1+K_1}(1+4\frac{{\lambda_1}}{\mu_1})$. The factor $\frac{2}{1+K_1}$ shows the effect of the penetration loss of $v_1$-orthogonal walls and $\frac{{\lambda_1}}{\mu_1}$ represents the effect of correlated shadowing by common obstacles. 
As both $\lambda_1$ and $\mu_1$ can be arbitrarily configured, \eqref{eq:compa_var_corr_uncorr} shows that the two models can yield arbitrarily different variances, highlighting the importance of including correlation.

\section{Success and Coverage Probabilities}\label{sec:Success and Coverage Probability}
In this section, we derive the coverage probability under the cellular network scenario and the success probability under the D2D underlay scenario, both in Poisson grid networks. The main assumption is that all BSs (and also all D2D links, in the D2D underlay scenario) share the same spectrum. In the cellular network setting, the typical user associates with one of the BSs and the other BSs act as interferers. In the D2D underlay setting, all signals from BSs are considered interference. 

For the D2D underlay scenario, we can directly apply the results on interference statistics obtained in the previous sections. For the case of cellular networks, the signal from the serving BS should be excluded from interference. We will first consider the D2D underlay case and then the cellular network case under several association scenarios. We assume that all D2D transmit powers are equal to ${P_{tx}}/{\nu}$ (i.e., $P_{link}=1/{\nu}$).

\subsection{Success Probability}\label{subsec:success_prob}
\subsubsection{Success Probability for a Single D2D Link}
We use the labeling system of Section \ref{sec:typical_room}.
\proposition\label{prop:succ_single_line} Under Rayleigh fading, the success probability of a D2D transmission from room $(i_1,i_2,\ldots,i_n)$ to room $(0,0,\ldots,0)$ is
{	
\begin{align*}
\textstyle\mathbb{P}[{\sf SINR}_s>\theta]=\mathcal{L}_{\tilde{I}_{(0,0,,\ldots,0)}}\left(\frac{\nu\theta}{\prod_{j=1}^nK_j^{|i_j|}}\right)\exp\left(-\frac{\nu\theta\sigma^2}{\prod_{j=1}^nK_j^{|i_j|}}\right),
\end{align*}}where $\mathcal{L}_{\tilde{I}_{(0,0,\ldots,0)}}(\cdot)$ is given in Proposition \ref{prop:laplace_dist_fading_lines}, $\theta$ is the SINR threshold and $\sigma^2$ is the thermal noise power.\begin{IEEEproof}
	Since the path loss model is (\ref{eq:path_loss_model}), the interference power from room $(i_1,i_2,\ldots,i_n)$ to the typical room is $h\prod_{j=1}^n K_j^{|i_j|}$, where $h$ is an exponential random variable with mean 1, the success probability is
	\begin{eqnarray*}
    \textstyle\mathbb{P}\left[\frac{h\prod_{j=1}^n K_j^{|i_j|}/\nu}{\tilde{I}_{(0,0,\ldots,0)}+\sigma^2}>\theta\right]=\mathbb{E}\exp\left(\frac{-\nu\theta(\tilde{I}_{(0,0,\ldots,0)}+\sigma^2)}{\prod_{j=1}^n K_j^{|i_j|}}\right)=\mathcal{L}_{\tilde{I}_{(0,0,,\ldots,0)}}\left(\frac{\nu\theta}{\prod_{j=1}^nK_j^{|i_j|}}\right)\exp\left(-\frac{\nu\theta\sigma^2}{\prod_{j=1}^nK_j^{|i_j|}}\right).\end{eqnarray*}\end{IEEEproof}

\begin{figure*}[t]
	\begin{minipage}{0.49\linewidth}
		\begin{center}
			\epsfxsize=2.8in {\epsfbox{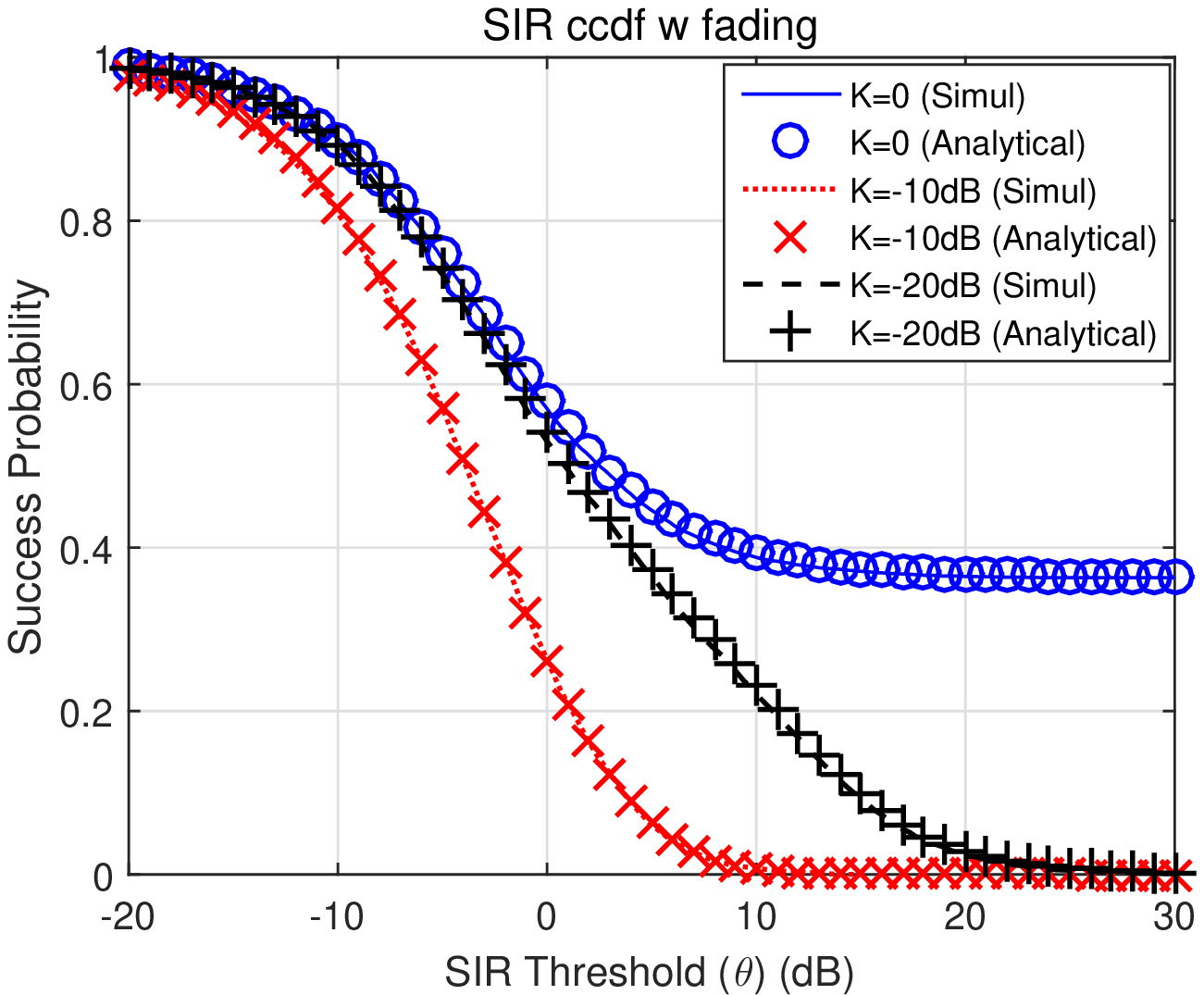}}
			\caption{\small In-room link success probability (D2D transmission within room ($0,0,0$)) under  the 3-D case, with $\nu=1$, $r_i=\frac{\lambda_i}{\mu_i} = 0.1$, $K_i=K$ for $i=1,2,3$ and $\sigma^2 = 0$.  }
			\label{fig:succ_prob_theta}
		\end{center}
	\end{minipage}
	\hspace{0.01\linewidth}
	\begin{minipage}{0.49\linewidth}
		\begin{center}
			\epsfxsize=2.8in {\epsfbox{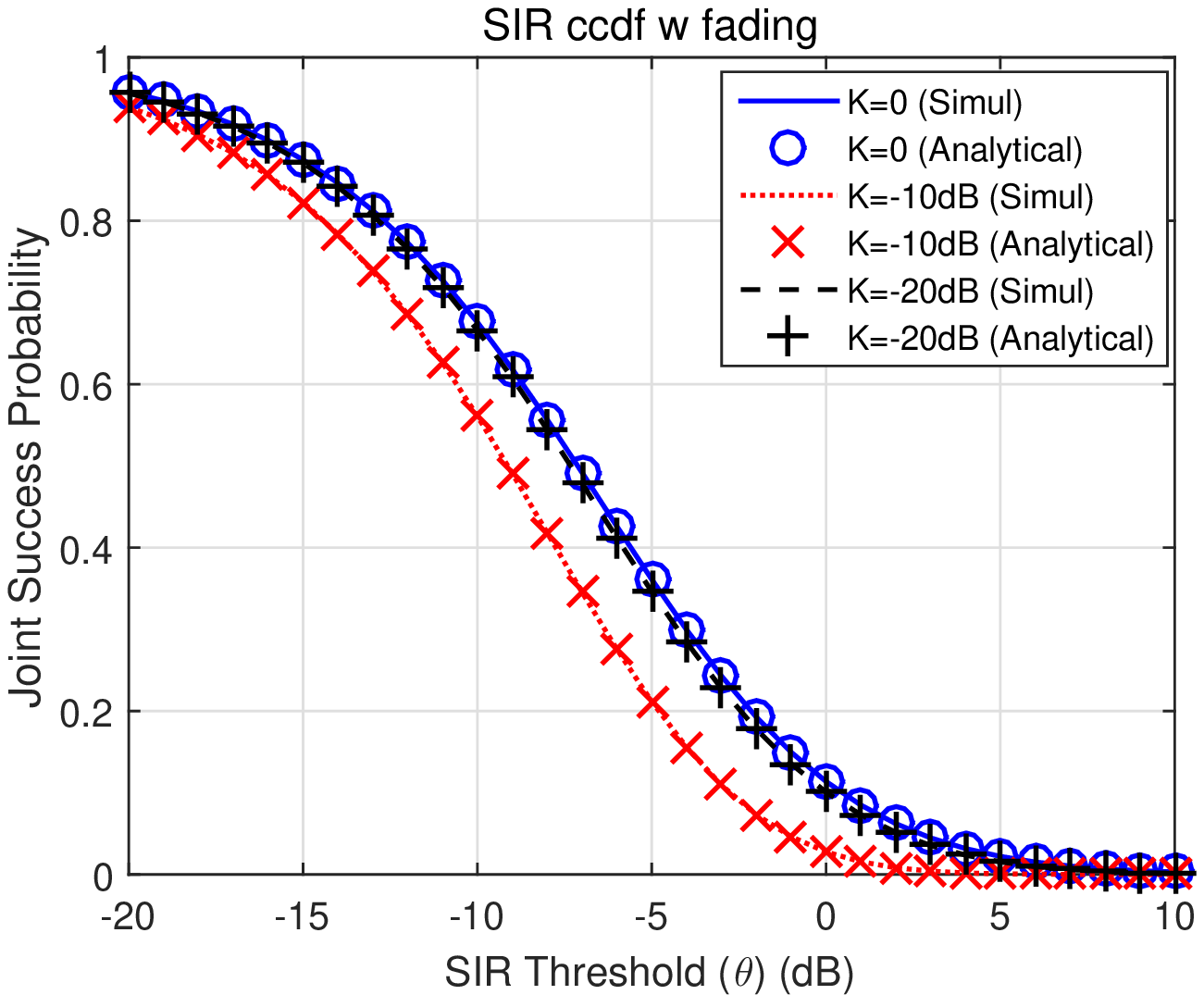}}
			\caption{\small In-room link joint success probability (Two D2D links attempt from ($0,0,0$) to itself) under  the 3-D case where $\nu=1$, $\theta=\theta'$, $r_i=\frac{\lambda_i}{\mu_i} = 0.1$, $K_i=K$ for $i=1,2,3$ and $\sigma^2_1,\sigma^2_2 = 0$.   
			}
			\label{fig:joint_succ_prob_theta}
		\end{center}
	\end{minipage}
\end{figure*}
Fig. \ref{fig:succ_prob_theta} illustrates the in-room link success probability when the both transceivers are in $(0,0,0)$ in the 3-D case. We assume $\nu=1$, $K_i=K$, $r_i=0.1$ for $i=1,2,3$ and $\sigma^2=0$. As $K$ decreases from $-10$dB to 0 ($-\infty$ dB), the interference power from other rooms is more attenuated and the success probability increases. When $K=0$, the success probability does not converge to 0 as $\theta$ tends to $\infty$ but converges to the probability that there exists no BS in room $(0,0,0)$.

\subsubsection{Success Probability for Multiple D2D Links}
We consider multiple D2D communications with normalized transmit powers. For a simple case, we assume there are two D2D links and these links are in-room links (i.e., a pair of TX and RX are in same room).
\proposition\label{prop:joint_success_prob} Consider two in-room D2D links, one in $(0,0,\ldots,0)$ and the other in $(i_1,i_2,\ldots,i_n)$, and $(i_1,i_2,\ldots,i_n)\neq(0,0,\ldots,0)$. Under Rayleigh fading, the probability that the SINRs of two D2D links are larger than $\theta$ and $\theta'$, respectively, is {
\begin{align*}
\textstyle\mathbb{P}[{\sf SINR}_1>\theta,{\sf SINR}_2>\theta']&\textstyle=\mathcal{L}_{\tilde{I}_{(0,0,\ldots,0)}\tilde{I}_{(i_1,i_2,\ldots,i_n)}}(\nu\theta,\nu\theta')\frac{1}{1+\nu\theta \prod_{m=1}^nK_m^{|i_m|}}\frac{1}{1+\nu\theta' \prod_{m=1}^nK_m^{|i_m|}}\\
&\textstyle\times\exp(-\nu(\theta\sigma_1^2+\theta'\sigma_2^2)),
\end{align*}}where $\mathcal{L}_{\tilde{I}_{(0,0,\ldots,0)}\tilde{I}_{(i_1,i_2,\ldots,i_n)}}(\cdot,\cdot)$ is given in Proposition \ref{prop:joint_laplace_transform}, $\sigma_1^2$ and $\sigma_2^2$ are the thermal noise powers of the first and the second link respectively. Here, SINR$_1$ and SINR$_2$ are the SINRs of the two D2D links.
The first term represents the interference from the BSs, the second and third terms the interference between the two D2D links, and the last one the thermal noise, respectively.

In Fig. \ref{fig:joint_succ_prob_theta}, the joint in-room success probability of two D2D links is plotted in the 3-D case. 
As in Fig. \ref{fig:succ_prob_theta}, the joint in-room success probability also increases as $K$ goes from $-10dB$ to 0.

\begin{figure*}[t]
	\begin{minipage}{0.31\linewidth}
		\begin{center}
			\epsfxsize=2.2in {\epsfbox{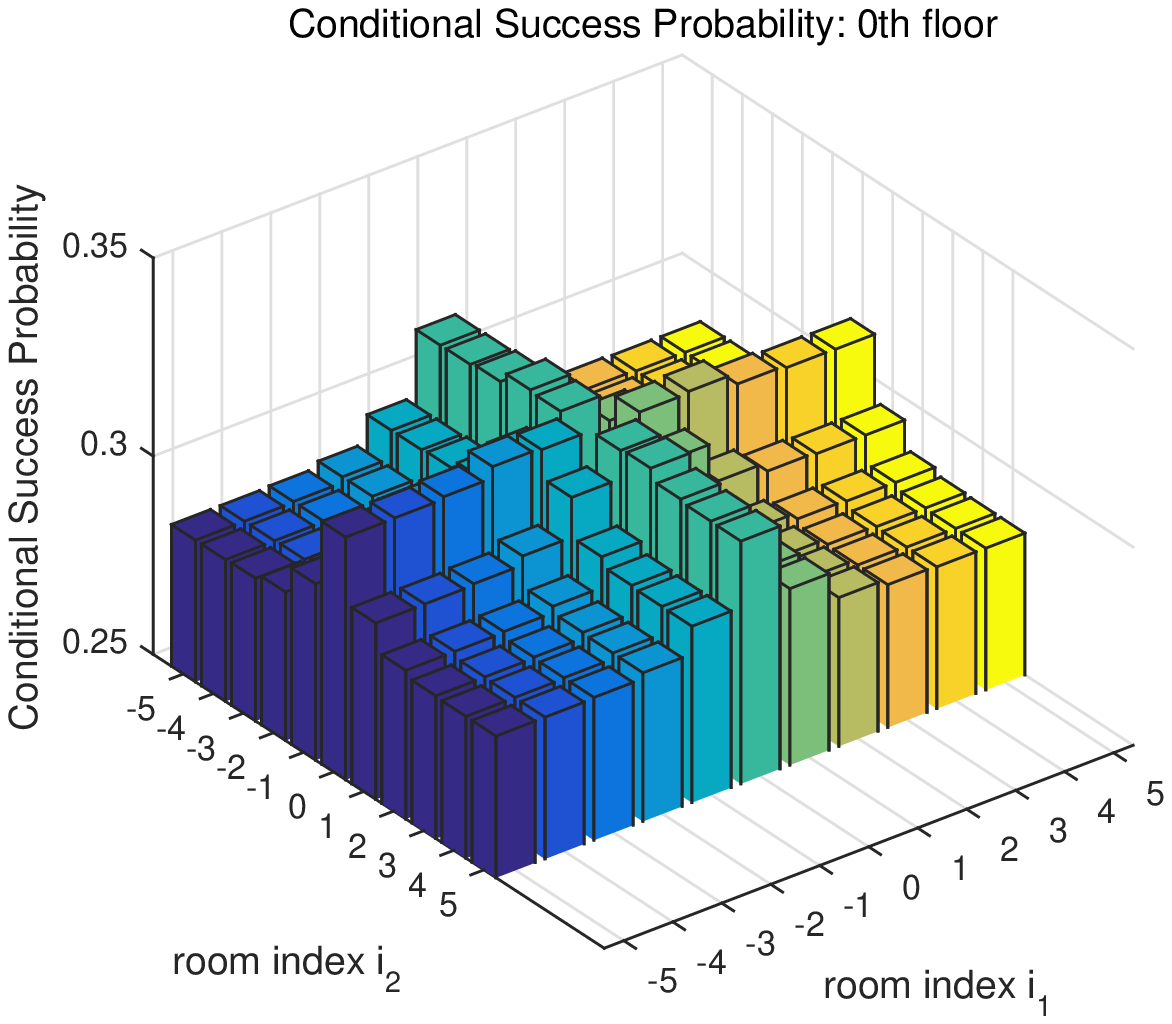}}
		\end{center}
	\end{minipage}
	\hspace{0.01\linewidth}
	\begin{minipage}{0.31\linewidth}
		\begin{center}
			\epsfxsize=2.2in {\epsfbox{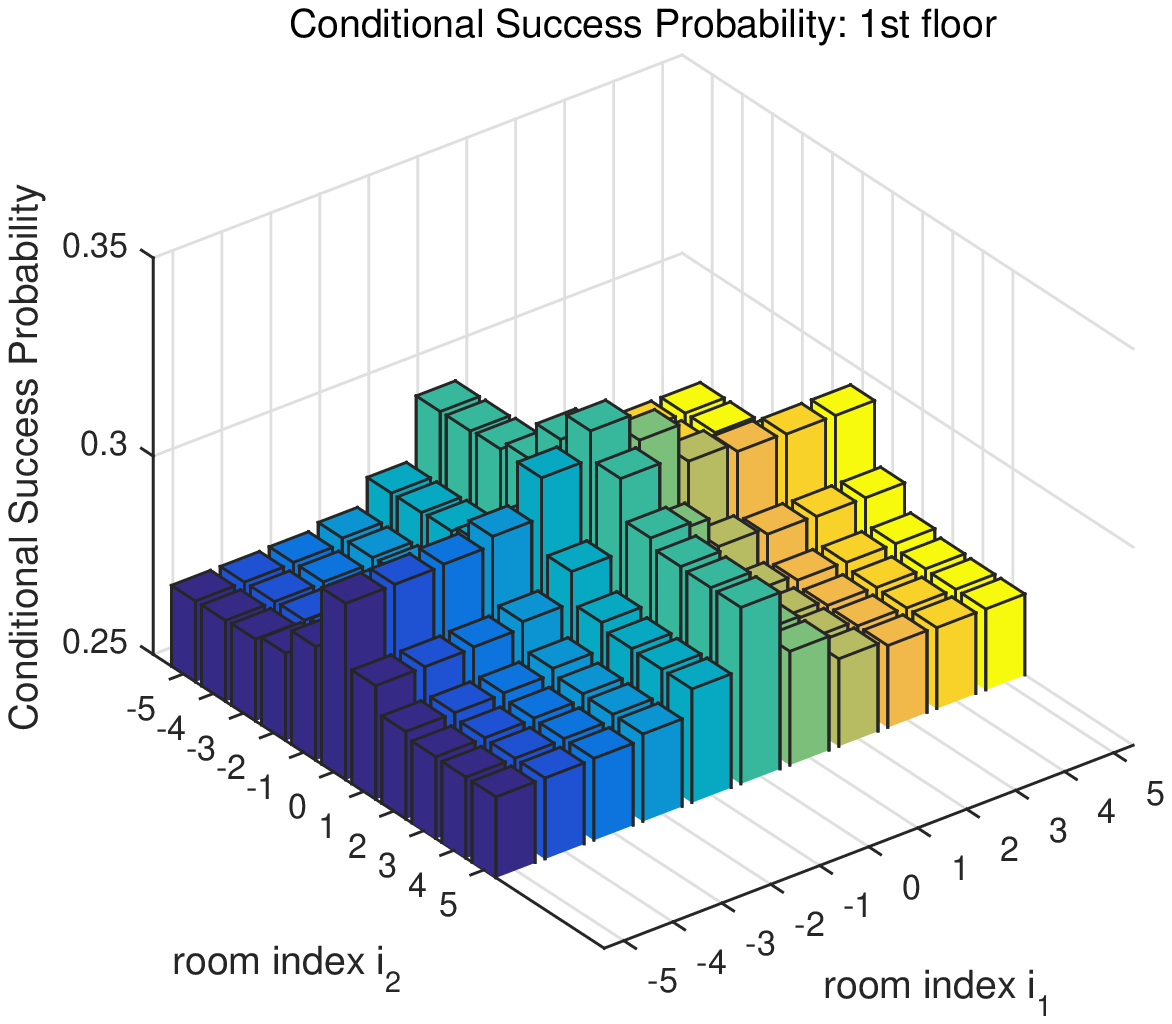}}
		\end{center}
	\end{minipage}
	\hspace{0.01\linewidth}
	\begin{minipage}{0.31\linewidth}
		\begin{center}
			\epsfxsize=2.2in {\epsfbox{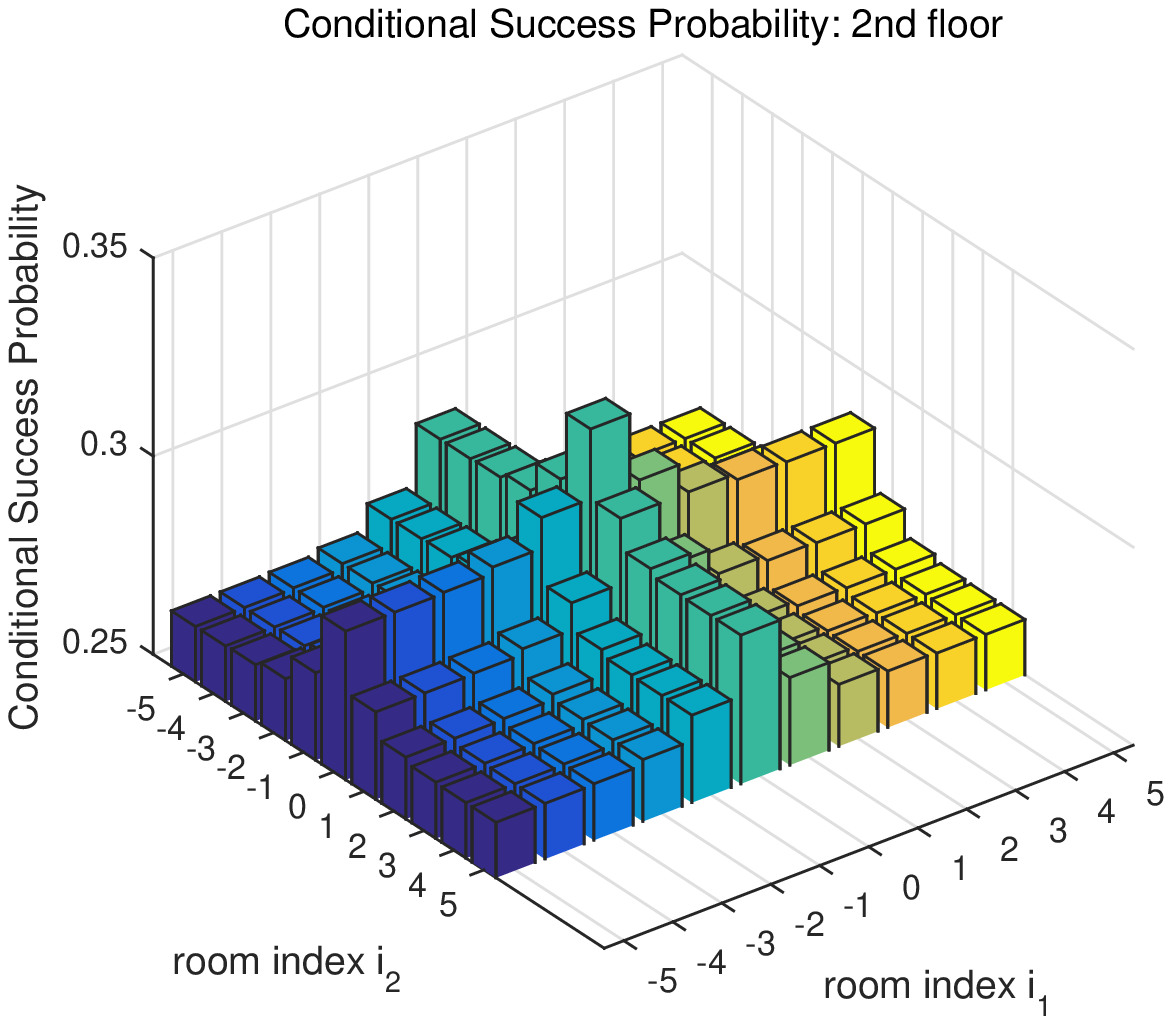}}
		\end{center}
	\end{minipage}
	\caption{\small Conditional probability that the SINR of a D2D link in room $(i_1,i_2,i_3)$ is larger than $0dB$ given the SINR of a D2D link in the typical room is larger than $0dB$ when $K_1,K_2,K_3 = -10dB$, $r_1,r_2,r_3=0.1$, and $\sigma_1^2,\sigma_2^2=0$.}
	\label{fig:cond_suc_prob}
\end{figure*}

Fig. \ref{fig:cond_suc_prob} illustrates the conditional probability that the second D2D link (in room ($i_1,i_2,i_3$)) is successful ($\mbox{SINR}_2>0dB$) given the first D2D link (in the typical room) is successful ($\mbox{SINR}_1>0dB$). 
The conditional success probability decreases as the room distance increase but in a non-isotropic manner, and the same room distance does not necessarily imply the same conditional success probability due to the intricate spatial interference correlation.\footnote{If two D2D links are in the same room (0,0,0), this leads to severe interference between these two links. Due to the randomness of channel fading coefficient, $\mathbb{P}[\mbox{SINR}_1>0dB]$ does not guarantee the highest $\mathbb{P}[\mbox{SINR}_2>0dB]$ when the second link is in the room $(0,0,0)$.}

\subsection{Coverage Probability}\label{subsec:coverage_prob}
In this part, we use the labeling system of Section \ref{sec:typical_user}. We consider two scenarios: One is the strongest BS association and the other is the nearest (graph-distance) BS association. If all $K_i$ are the same and there is no fading ($h\equiv 1$), these two scenarios are the same. 
With fading, the nearest BS is not always the strongest one. 
\subsubsection{Strongest BS Association} First, we consider the case where the typical user associates to the BS which provides the best (strongest) instantaneous signal.

\proposition\label{prop:strong} Under Rayleigh fading and strongest BS association, the coverage probability is { 
	\begin{align}\label{eq:strong_cov}
	\textstyle\sum_{k=1}^{n}p(\theta,\lambda_i,\left(K_{(i+k)\%n}\right)_{i=0}^{n-1},2^{n-1}\left(r_{(i+k)\%n}\right)_{i=0}^{n-1}),
	\end{align}
}where $\theta> 1$ and  {	\begin{align*}
	\scriptstyle& \scriptstyle p(\theta,\lambda,\left(K_{(i)\%n}\right)_{i=0}^{n-1},\left(r_{(i)\%n}\right)_{i=0}^{n-1})=p(\theta,\lambda,K_1,K_2,\ldots,K_n,r_1,r_2,\ldots,r_n)\\
	\scriptstyle = & \scriptstyle 2^n\lambda\sum_{j_1\in\mathbb{N}}\sum_{(j_t)_{t=2}^n\in\mathbb{Z}^{n-1}}\exp(-\frac{K_1\theta \sigma^2}{\prod_{m=1}^nK_m^{|j_m|}})\left({1+r_i\sum_{(l_t)_{t=2}^n\in\mathbb{Z}^{n-1}}\frac{1}{1+\prod_{m=2}^n K_m^{|j_m|-|l_m|}/\theta }}\right)^{-1}\nonumber\\
\scriptstyle 	&\scriptstyle \times  \prod_{m=1}^n\left(\prod_{l_m\in\mathbb{N}} \left({1+r_m\sum_{(l_t)_{t=1,\ldots,m}^{\neq m}\in\mathbb{Z}^{n-1}}\frac{1/K_m}{1/K_m +K_1^{|j_1|-|l_1|-1}\prod_{m=2}^nK_m^{|j_m|-|l_m|}/\theta }}          \right)^{-2}\right). 
	\end{align*}
}Here $\sigma^2$ is the thermal noise power. If $\theta \leq 1$, the
equality in (\ref{eq:strong_cov}) should be replaced by $\leq$.\begin{IEEEproof}[Proof (sketch)]
  For any BS $t\in\Phi$, let $N_1(t),N_2(t),\ldots,N_n(t)$ be the number of hyperplanes between $t$ and the typical user. Let $d\in\Phi$ be the serving transmitter. The aggregated received power from all transmitters except $d$ is{
	\begin{align*}
\textstyle	I^{!d}=\sum_{t\in\Phi\setminus\{d\}}h_t \prod_{m=1}^nK_m^{N_m(t)},
	\end{align*}}where $h_t$ is the channel fading coefficient between transmitter $t$ and the typical user. Then, the coverage probability is
{	
	\begin{eqnarray*}
	\textstyle\mathbb{P}[{\sf SINR}_c>\theta]&\textstyle=&\textstyle\mathbb{P}(\text{max}\frac{h_d\prod_{m=1}^nK_m^{N_m}}{I^{!d}+\sigma^2}>\theta)=\mathbb{E}\mathds{1}(\text{max}\frac{h_d\prod_{m=1}^nK_m^{N_m}}{I^{!d}+\sigma^2}>\theta)\\
	\textstyle&\textstyle\stackrel{(a)}{\leq}&\textstyle\mathbb{E}\sum_{d\in\Phi}\mathds{1}(\frac{h_d\prod_{m=1}^nK_m^{N_m}}{I^{!d}+\sigma^2}>\theta)\\
	\textstyle&\textstyle=&\textstyle\mathbb{E}[\sum_{d\in\Phi\cup \bar{v}_1}\mathds{1}(\frac{h_d\prod_{m=1}^nK_m^{N_m}}{I^{!d}+\sigma^2}>\theta)+\ldots+\sum_{d\in\Phi\cup \bar{v}_n}\mathds{1}(\frac{h_d\prod_{m=1}^nK_m^{N_m}}{I^{!d}+\sigma^2}>\theta)],
	\end{eqnarray*}}where $h_d$ is the channel coefficient between $d$ and the typical user, and $\bar{v}_i, i\in[n]$ are the line segments parallel to the $v_i$ axis. $(a)$ comes from the fact that if $\theta> 1$, there is (almost surely) at most one BS serving ${\sf SINR}>\theta$\cite{dhillon2012modeling}. If we only consider the BSs on $v_1$ in the given structure $\Psi$, the conditioned expression becomes{
	\begin{eqnarray*}
		\textstyle\mathbb{E}[\sum_{d\in\Phi\cup \bar{v}_1}\mathds{1}(\frac{h_d\prod_{m=1}^nK_m^{N_m}}{I^{!d}+\sigma^2}>\theta)|\Psi]
		=2^{n-1}\lambda_1 \sum_{(j_t)_{t=1}^n\in\mathbb{Z}^n}^{j_1\neq 0}l_{1_i}\mathbb{P}(\frac{h_d\prod_{m=1}^nK_m^{N_m}}{K_1(I^{!d}+\sigma^2)}>\theta),
	\end{eqnarray*}}by Slivnyak's theorem\cite{baccelli2009stochastic,haenggi2012stochastic} and the construction of the Poisson grid. We obtain the result by deconditioning with respect to the Poisson grid, the channel coefficients.
\end{IEEEproof}

\begin{figure*}[t]
	\begin{minipage}{0.49\linewidth}
		\begin{center}
			\epsfxsize=2.8in {\epsfbox{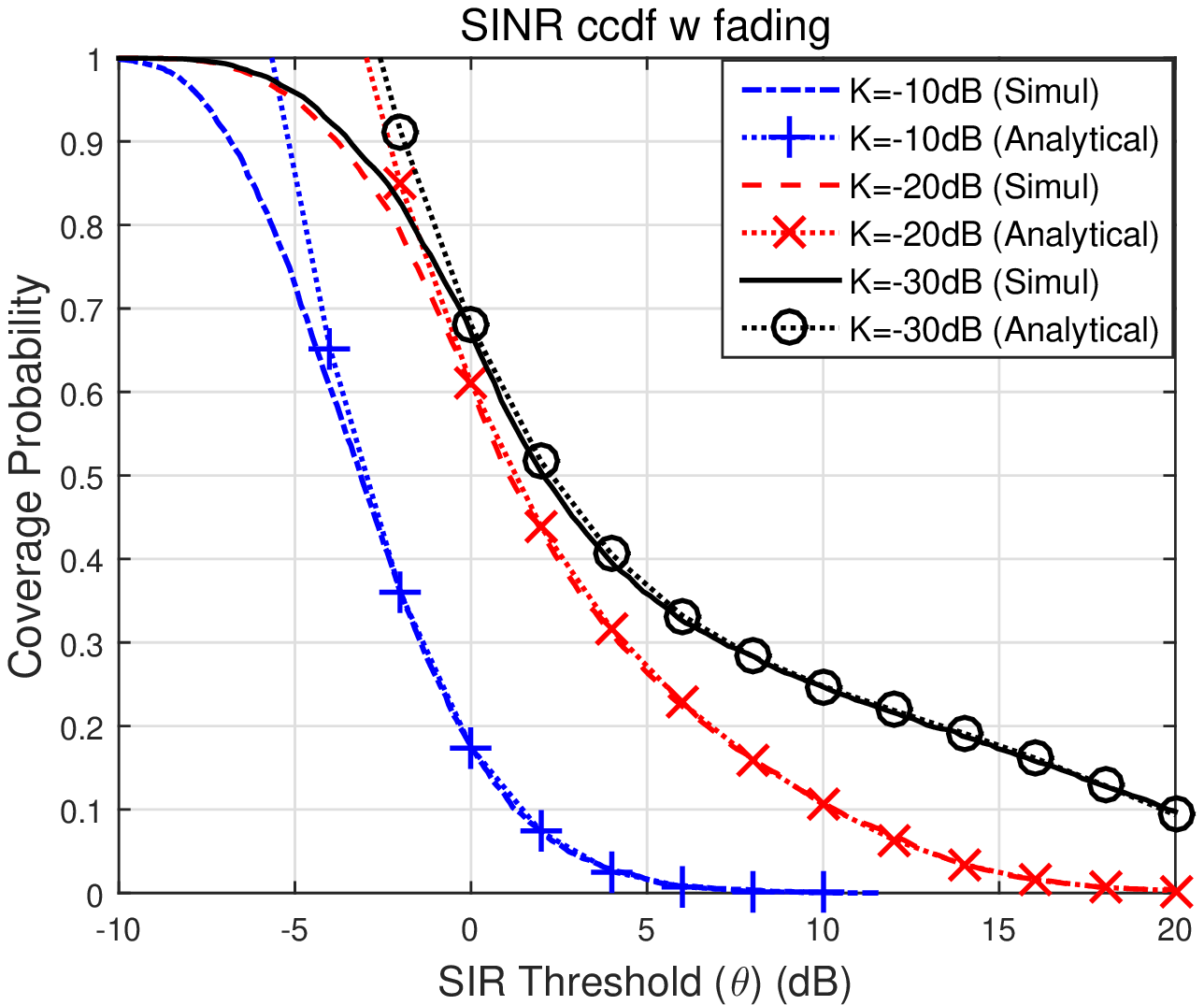}}
			\caption{\small Coverage Probability under the 3-D case and the strongest association scenario. Here $K_i = K$, and $r_i=\frac{\lambda_i}{\mu_i} = 0.1$ for $i=1,2,3$ and $\sigma^2=0$  }
			\label{fig:cov_prob_strong}
		\end{center}
	\end{minipage}
	\hspace{0.01\linewidth}
	\begin{minipage}{0.49\linewidth}
		\begin{center}
			\epsfxsize=2.8in {\epsfbox{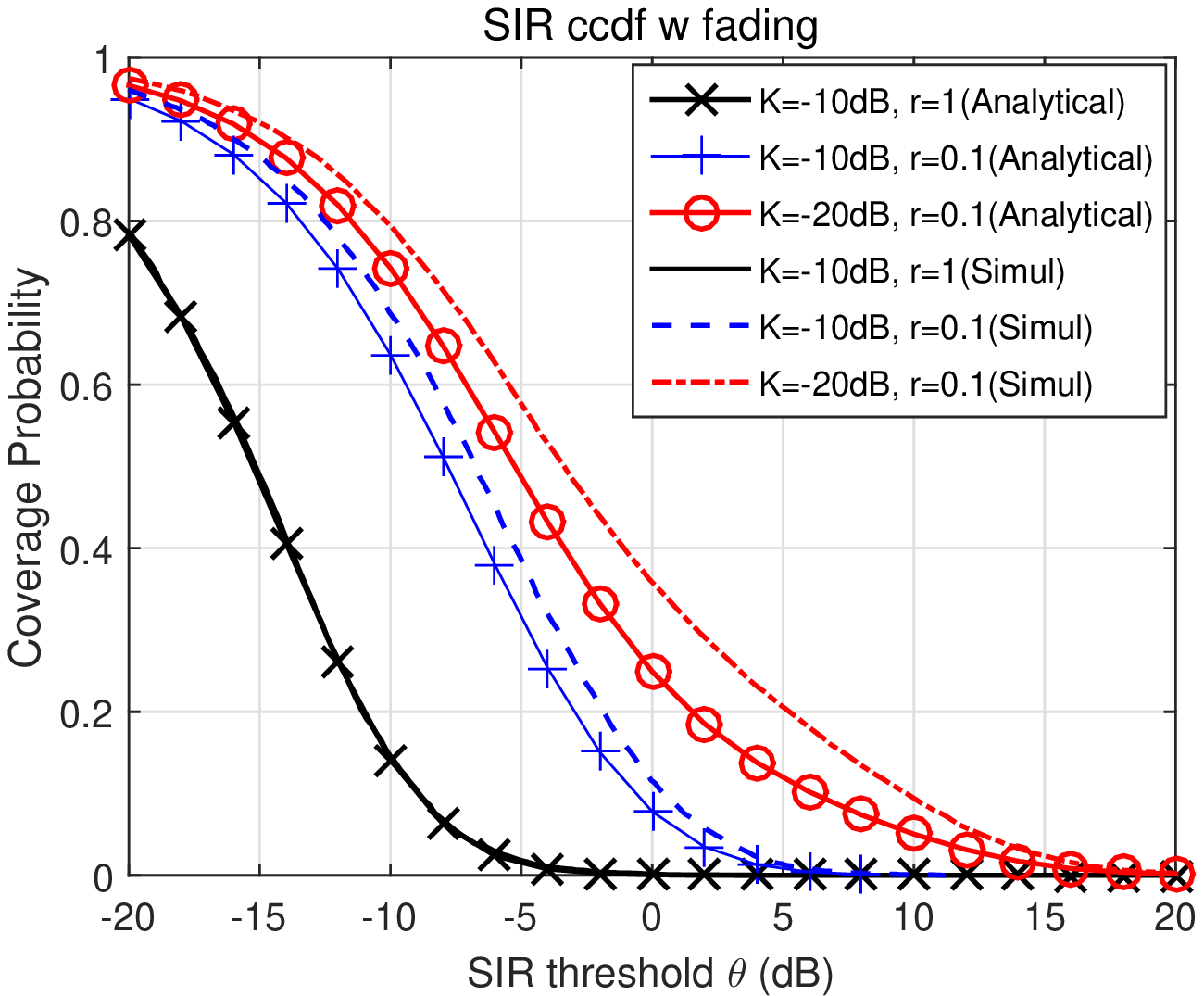}}
			\caption{\small Coverage Probability under the 3-D case and the nearest association scenario. $K_i = K$, and $r_i=\frac{\lambda_i}{\mu_i} = r$ for $i=1,2,3$ and $\sigma^2=0$  
			}
			\label{fig:cov_prob_near}
		\end{center}
	\end{minipage}
\end{figure*}

\subsubsection{Nearest (Room-Distance) BS Association} In the case where the user connects to the BS with strongest average signal, the BS selection process boils down to finding the nearest (w.r.t. room-distance) BS. More precisely, let $K_i=K, i\in[n]$, and define the graph distance between two rooms as the number of hyperplanes between these rooms. The typical user is associated to the minimal graph distance BS. Denote this distance by $\delta$. If there are several BSs at distance $\delta$, the user associates randomly with one of them.

\remark  Given the Poisson grid $\Psi$ and $\delta$, one can compute the coverage probability in closed form. In the 3-D case, assume $\delta=0$ and define $N$ to be the number of BSs in the tagged room (or $(0,0,0)$). Then,{
\begin{align*}
\textstyle\mathbb{P}[N=1|\delta=0,\Psi]=\sum_{i=1}^{12}\lambda_i\times\exp(-\sum_{i=1}^{12}\lambda_i)/P_{tot},
\end{align*}}where $\lambda_i,i\in[1,12]$ is the mean number of BSs on the edges of the tagged room and $P_{tot}=\mathbb{P}[\delta=0|\Psi]$. For the cases $N=2,3,\ldots$, it is possible to compute these quantities. So, conditional on the given Poisson grid $\Psi$, the Laplace transform $\mathbb{E}[e^{-sI}|\Psi,\delta=0]$ can be computed using Proposition \ref{prop:laplace_dist_fading_lines_user} and $\mathbb{P}[N=n|\Psi,\delta=0]$. The conditional interference Laplace transform under these conditions is given in Appendix \ref{appendix:nearest_conditioned}.

\proposition Under Rayleigh fading, the coverage probability of a typical user, $\mathbb{P}[{\sf SINR}_c>\theta]$ is asymptotically equal to{
\begin{align}\label{eq:strong}
\textstyle\sum_{m\geq 0}\exp(-\frac{\theta\sigma^2}{K^m})\times\left(\prod_{i=1}^mh_1(m,2^{n-1}r_i,\frac{\theta}{K^m})^2-\prod_{i=1}^mh_2(m,2^{n-1}r_i,\frac{\theta}{K^m})^2\right),
\end{align} }as $r_i$ goes to infinity, where $\sigma^2$ is the thermal noise power and 
{
\begin{align}\label{eq:nearest_eq_part}
 \scriptstyle h_1(m,r,s) &\scriptstyle = \left(\right.\prod_{i_1\in\mathbb{N}}1+r((1+2^{m-1}(m+1-i_1)(m-i_1))V_m(i_1)
+\sum_{(i_t)_{t=2}^{n}\in\mathbb{Z}^{n-1}}^{i_1+\sum_{t=2}^n|i_t|\geq n+1}(1-\frac{K}{K+sK^{i_1+\sum_{t=2}^n|i_t|}}))\left.\right)^{-1}\nonumber\\
 \scriptstyle h_2(m,r,s) &\scriptstyle = \left(\right.\prod_{i_1\in\mathbb{N}}1+r((1+2^{n-1}(m+2-i_1)(m+1-i_1))V_{m+1}(i_1)+\sum_{(i_t)_{t=2}^{n}\in\mathbb{Z}^{n-1}}^{i_1+\sum_{t=2}^n|i_t|\geq n+1}(1-\frac{K}{K+sK^{i_1+\sum_{t=2}^n|i_t|}}))\left.\right)^{-1},\nonumber
\end{align}}and $V_m(x)$ is 1 if $x=1,2,\ldots,m$ and $0$ otherwise. When $r_i, i\in[n]$ goes to infinity, this lower bound is asymptotically tight.\begin{IEEEproof}[Proof (sketch)]
	Assume we have at least one BS on a specific line segment. Denote the number of BSs on that line by $N$ and the density by $\lambda$. Then, $\mathbb{P}(N=n+1|N>0)=e^{-\lambda}\frac{\lambda^{n+1}}{(n+1)!(1-e^{-\lambda})}$. As $\lambda$ increases, this distribution converges to $\mathbb{P}(N=n+1)$, and $\mathbb{P}(N=0)$ goes to $0$. With this result and Proposition \ref{prop:laplace_dist_fading_lines_user}, we can obtain the asymptotic conditional coverage probability given the Poisson grid, and then obtain (\ref{eq:strong}) by deconditioning.
\end{IEEEproof}

Fig. \ref{fig:cov_prob_near} describes the coverage probability under the nearest room-distance BS association scenario in the 3-D case. We assume that $K_i=K$, $r_i=r$ for $i=1,2,3$ and $\sigma^2=0$. As $r$ increases, the gap between (\ref{eq:strong}) and the simulation result decreases. Fig. \ref{fig:cov_prob_near} shows that our expression (\ref{eq:strong}) matches well with the simulation result when $r > 0.1$.

\remark If we compare the case of $K=-10dB, r=0.1$ in Figs. \ref{fig:cov_prob_strong} and \ref{fig:cov_prob_near}, the coverage probability under the strongest BS association case is higher than the nearest room distance BS association case, which is inline with intuition. The difference between these two cases provides quantitative guidelines for determining the worthiness of pursuing instantaneous cell reselection.

\subsection{3-D Free-space and Poisson Building}\label{subsec:two_model_comparison}

\begin{figure*}[t]
	\begin{minipage}{0.49\linewidth}
		\begin{center}
		\epsfxsize=2.8in {\epsfbox{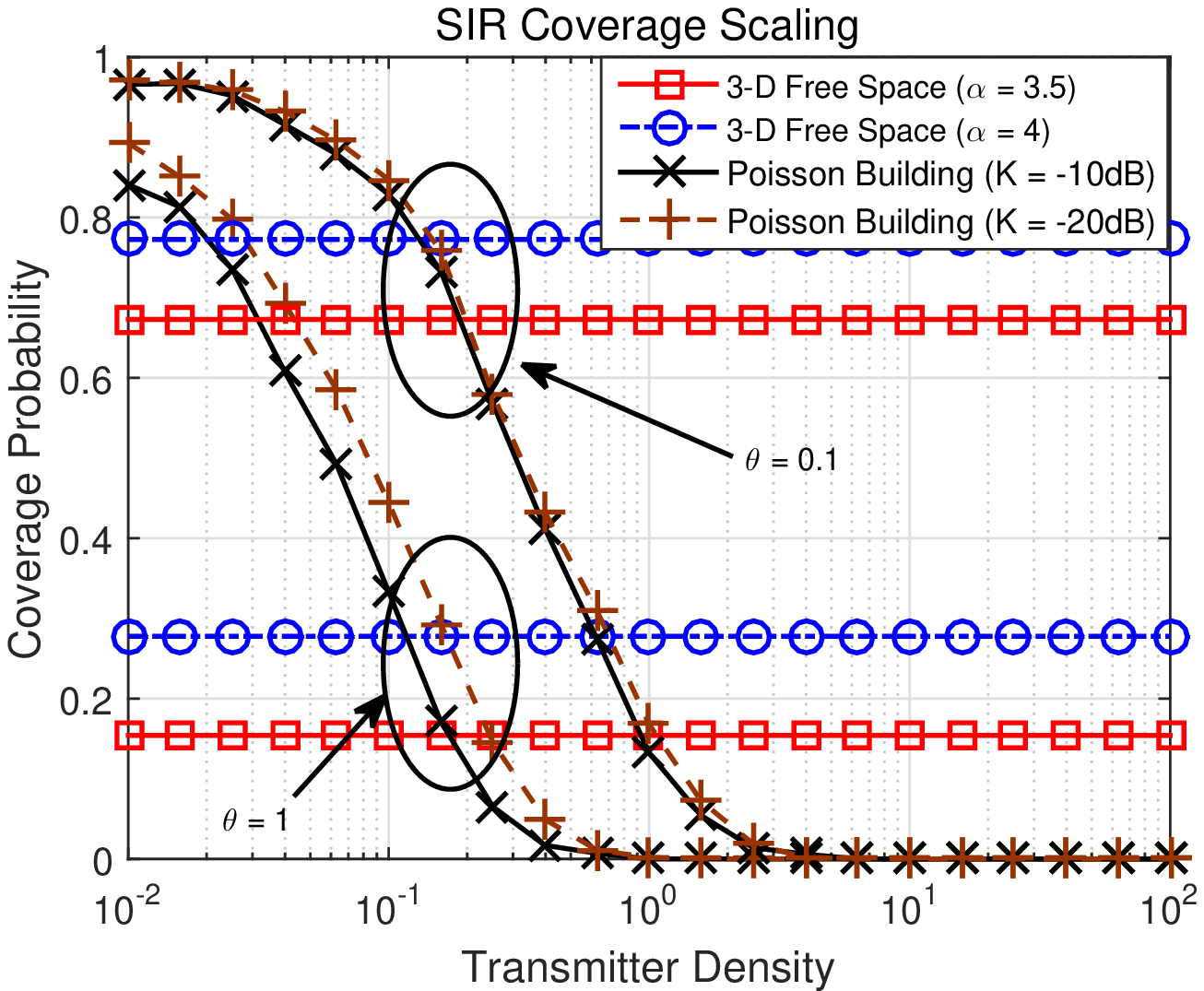}}
		\caption{\small SIR coverage scaling over network density   }
		\label{fig:SIR_scaling}
    \end{center}
  \end{minipage}
  \begin{minipage}{0.49\linewidth}
		\begin{center}
		\epsfxsize=2.8in {\epsfbox{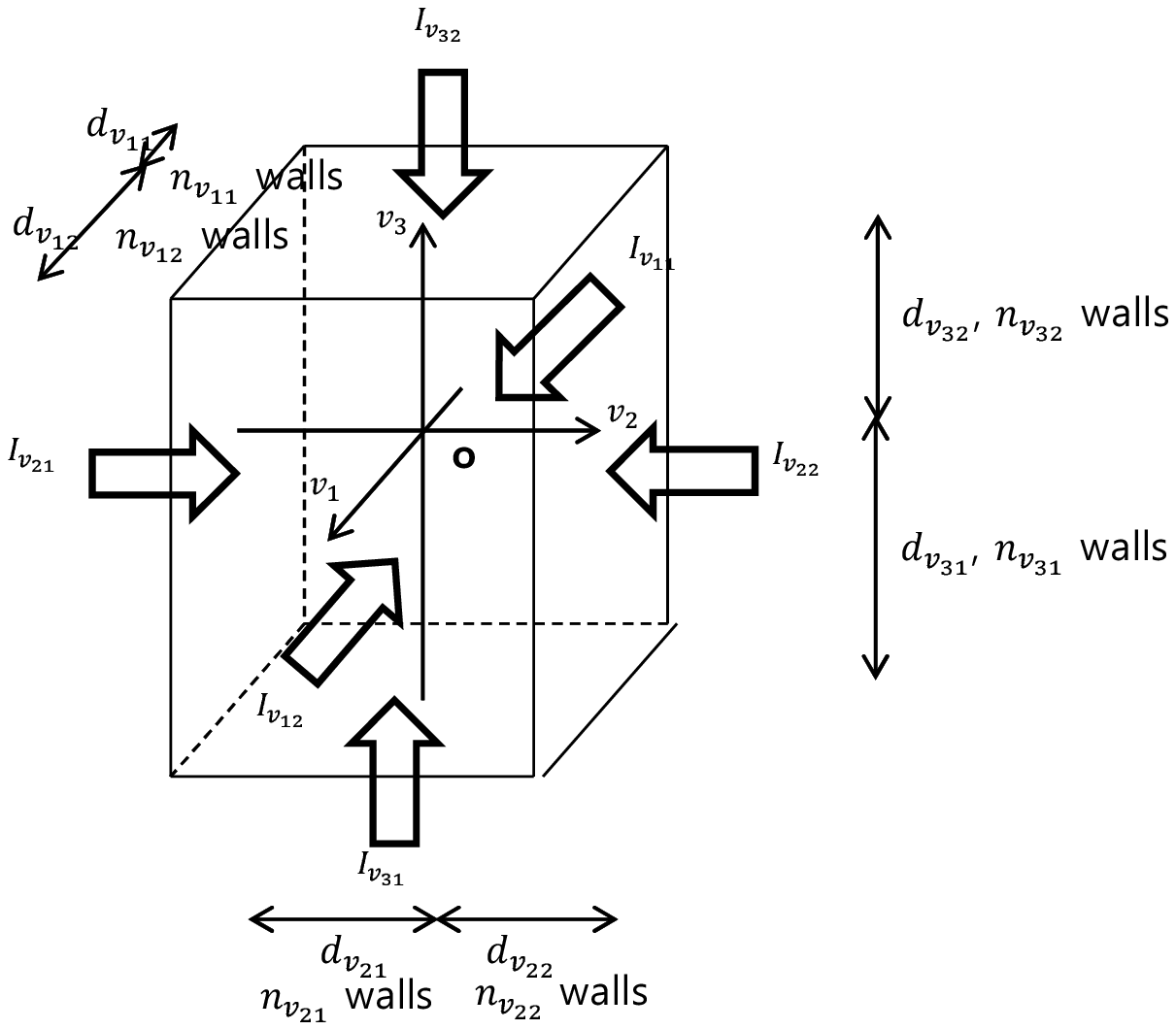}}
		\caption{\small Finite Size Poisson Building }
		\label{fig:finite_size}
    \end{center}
  \end{minipage}
\end{figure*}

In this paper, we suggest a new network model with the blockage-based path loss model. In this subsection, we provide a justification of our model for in-building networks, by comparing the SIR distribution of a classical stochastic geometric model without correlated shadowing and our new model.
Fig. \ref{fig:SIR_scaling} compares the 3-D free space model with distance-based path loss function and the Poisson building with blockage-based path loss function under the nearest BS (or nearest room-distance) association. The free space model is the 3-D extension of the network model in \cite{net:Andrews12tcom} with path loss exponents $\alpha=3.5,4$. The Poisson building is constructed with $\mu_i = 1$ and $K_i = -10 \mbox{,} -20dB$ for $i=1,2,3$. We use the average BS density (\ref{eq:cox_avg_lambda}), $\small \lambda_{avg}=\left(\frac{4\lambda_1}{\mu_1}+\frac{4\lambda_2}{\mu_2}+\frac{4\lambda_3}{\mu_3}\right)\mu_1\mu_2\mu_3$, the mean number of BSs per cubic meter in the Poisson building.

In the free-space model, the SIR scale invariance, \emph{i.e.,} the fact that the SIR at the typical user does not depend on the infrastructure density, which was observed in 2-D in \cite{ZhangAndrews2015,net:Andrews12tcom}, can be generalized in 3-D as shown in Fig. \ref{fig:SIR_scaling}, whereas under the \emph{Poisson building} model, the SIR coverage decreases rapidly with the average BS density. Clearly, 3-D in-building model cannot be reduced to a 3-D free-space model from the above observation.

\section{Finite Poisson Structures}\label{sec:finite_size}

The analysis so far focuses on an infinite network, which circumvents the boundary effects and thus brings extra tractability. While such a modeling approach is justifiable for low dimensionality (2-D) networks, which represent large cities, the boundary effects kick in much sooner in higher dimension (3-D).
Fortunately, the Poisson grid model can be tailored to analyze networks of finite sizes (with acceptable loss of tractability).
In this section, we analyze the interference observed by the typical user and compare it with the results Section~\ref{sec:typical_user}.

\subsection{Finite Size 3-D Poisson Building}
 We assume that the dimension of the building is $[-d_{v_{11}},d_{v_{12}}]\times [-d_{v_{21}},d_{v_{22}}]\times[-d_{v_{31}},d_{v_{32}}]\in\mathbb{R}^3$. As in Fig. \ref{fig:finite_size} denote by $n_{v_{11}}$ ($n_{v_{12}}$, $n_{v_{21}}$, $n_{v_{22}}$, $n_{v_{31}}$, $n_{v_{32}}$, resp.), the number of walls between the typical user and building boundary toward $-v_{1}$ ($+v_{1}$, $-v_{2}$, $+v_{2}$, $-v_{3}$, $+v_{3}$, resp.)-axis direction. Further, we denote by $I_{v_{11}}$, $I_{v_{12}}$, $I_{v_{21}}$, $I_{v_{22}}$, $I_{v_{31}}$, $I_{v_{32}}$, the interference coming from outside the finite building along the $-v_1$, $+v_1$, $-v_2$, $+v_2$, $-v_3$, $+v_3$, respectively.
\proposition\label{prop:finite_building} Under Rayleigh fading, the Laplace transform of the interference at the origin in a finite size Poisson building given $d_{v_{ij}}(i\in[1,2,3],j\in[1,2])$ and the out of building (OoB) interference $I_{v_{ij}}(i\in[1,2,3],j\in[1,2])$, is {\small $\mathbb{E}\left[e^{-sI}|d_{v_{ij}},I_{v_{ij}},(i\in[1,2,3],j\in[1,2])\right]=$}	\begin{eqnarray*}
	& &\textstyle \sum_{n_{v_{ij}}=1\ldots\infty}^{i=[1,2,3],j=[1,2]}\mathlarger{\mathlarger{\left[\right.}}\prod_{(i,j)}\frac{1}{1+sI_{v_{ij}}K_i^{n_{v_{ij}}}}l(\mu_i,d_{v_{ij}},\lambda_i,n_{v_{ij}},n_{v_{(i+1)\%3(j)\%2}}, \nonumber\\
	& &\textstyle n_{v_{(i+1)\%3(j+1)\%2}},n_{v_{(i+2)\%3(j)\%2}},n_{v_{(i+2)\%3(j+1)\%2}},K_i,K_{(i+1)\%3},K_{(i+2)\%3},s)
	\mathlarger{\mathlarger{\left.\right]}}\mbox{,}
\end{eqnarray*}where  
	\begin{eqnarray*}
		&&l(\mu,d,\lambda,n_{v_{11}},n_{v_{21}},n_{v_{22}},n_{v_{31}},n_{v_{32}},K_x,K_y,K_z,s)\\
		&&\textstyle=\frac{\mu^{n_{v_{11}}-1}e^{-\mu d}}{2\pi j}\int^{c+j\infty}_{c-j\infty}\exp(zd)\prod_{i=1}^{n_{v_{11}}}\frac{1}{z+{4\lambda}\sum_{j=-n_{v_{21}}+1}^{n_{v_{22}}-1}\sum_{k=-n_{v_{31}}+1}^{n_{v_{32}}-1}(1-\frac{1}{1+sK_1^{|i|-1}K_2^{|j|}K_3^{|k|}})}\text{dz}.
	\end{eqnarray*}	 \begin{IEEEproof}[Proof (sketch)]
	See Appendix \ref{appendix: finite_building}.
\end{IEEEproof}

\begin{figure}
	\begin{center}
		\epsfxsize=2.8in {\epsfbox{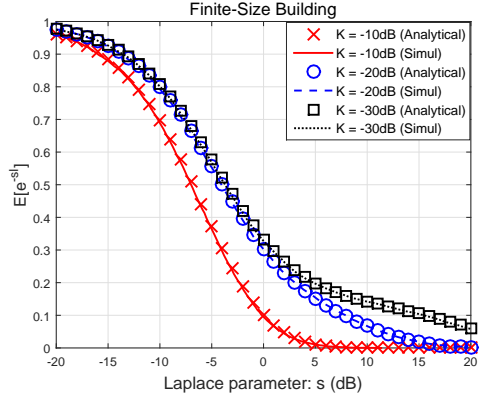}}
		\caption{\small Interference distribution in a Poisson building when $d_{v_{11}}=3$, $d_{v_{12}}=d_{v_{21}}=d_{v_{22}}=d_{v_{31}}=d_{v_{32}}=\infty$, $\lambda_i=0.1$, $\mu_i=1$, and $K_i=K$ for $i=1,2,3$. 
			}
		\label{fig:finite_size_simul}
	\end{center}
\end{figure}

\remark We consider the special case where $d_{v_{12}}$, $d_{v_{21}}$, $d_{v_{22}}$, $d_{v_{31}}$, $d_{v_{32}}$ $=\infty$, and there is no interference from outside. By Proposition \ref{prop:laplace_dist_fading_lines_user}, Proposition \ref{prop:finite_building} reduces to $\mathbb{E}\left[e^{-sI}|d_{x_1}\right]=${
\begin{eqnarray*}
	 \scriptstyle& \scriptstyle&\scriptstyle \sum_{n=1}^{\infty}\frac{\mu_1^{n-1}e^{-\mu d_{v_{11}}}}{2\pi j}\int_{c-j\infty}^{c+j\infty}\exp(zd_{v_{11}})\prod_{i=1}^n\left({z+{4\lambda_1}\sum_{j,k=-\infty}^{\infty}(1-\frac{K_1}{K_1+sK_1^{i}K_2^{|j|}K_3^{|k|}})}\right)^{-1}\\
	\scriptstyle&\scriptstyle\times&\scriptstyle\prod_{i=1}^{\infty}\left({1+\frac{4\lambda_1}{\mu_1}\sum_{j,k=-\infty}^{\infty}(1-\frac{K_1}{K_1+sK_1^{i}K_2^{|j|}K_3^{|k|}})}\right)^{-1}\\
	\scriptstyle&\scriptstyle\times&\scriptstyle\left(\prod_{j=1}^{\infty}{1+\frac{4\lambda_2}{\mu_2}\sum_{i,k=-\infty}^{\infty}(1-\frac{K_2}{K_2+sK_1^{|i|}K_2^{j}K_3^{|k|}})}\right)^{-2}\left(\prod_{k=1}^{\infty}{1+\frac{4\lambda_3}{\mu_3}\sum_{i,j=-\infty}^{\infty}(1-\frac{K_3}{K_3+sK_1^{|i|}K_2^{|j|}K_3^{k}})}\right)^{-2}.
\end{eqnarray*}}This is a simplified version of Proposition \ref{prop:finite_building} with elements from Proposition \ref{prop:laplace_dist_fading_lines_user}. 
Fig. \ref{fig:finite_size_simul} illustrates the Laplace transform of the interference observed by the typical user when $d_{v_{11}}=3$, $d_{v_{12}},d_{v_{21}},d_{v_{22}},d_{v_{31}},d_{v_{32}}=\infty$, $\lambda_1,\lambda_2,\lambda_3=0.1$, $\mu_1,\mu_2,\mu_3=1$, $K_1,K_2,K_3=K$.

\subsection{Window Office}
In a real environment, even if interference does not penetrate the floors (i.e., $K_3=0$), the interference from other floors can enter a room through paths outside the building, \emph{e.g.,} by reflecting on the neighboring buildings. To analyze this type of interference, we propose a semi-infinite Poisson building which is only deployed on the positive half plane of the $v_1$-axis. This building has a boundary wall at $v_1=0$ and the rooms with this boundary wall will be referred as window rooms.

\begin{figure}
	\begin{center}
		\epsfxsize=3.5in {\epsfbox{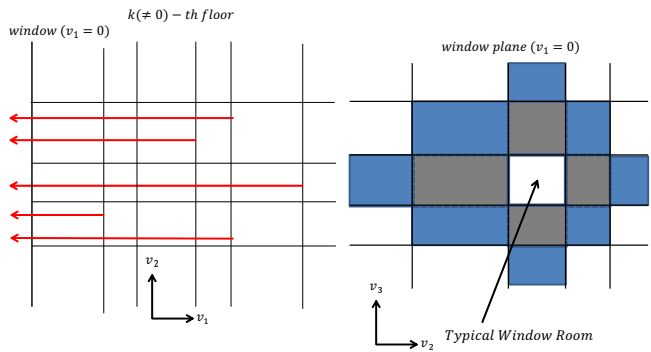}}
		\caption{\small Semi-infinite Poisson Building with the signal paths from a transmitter in room $(i,j,k)$ to the typical window room $(0,0,0)$. The left figure describes that signal from a transmitter passes to the outside through the shortest path. The right one describes the graph-distance based path loss model used in the free-space region.}
		\label{fig:window_office}
	\end{center}
\end{figure}

We choose a window room which contains the origin of the 3-D Euclidean space and label this room as $(0,0,0)$ (this is the typical room perspective as in Section \ref{sec:typical_room}), and label the other rooms according to their relative position with respect to room $(0,0,0)$. We only consider (shortest) graph-distance paths of signals to the outside\footnote{This is justified by the dominance of penetration loss over free-space loss in indoor.}. For example, if a BS is in room $(i,j,k)$, then a signal from this transmitter predominantly emits to outside through a window room $(0,j,k)$. See the left figure in Fig. \ref{fig:window_office}. Since room sizes in our Poisson building model are random, here we apply a hypothetical path loss function. We define the path loss from a window room $(0,j,k)$ to $(0,0,0)$ through OoB paths to be $l_{|j|+|k|}\in[0,1]$\footnote{Since the Poisson grid is a discrete random structure, it is hard to combine distance-based functions with it. Instead of this, we take a general path loss level-set function using graph distance as indicated in the right of Fig. \ref{fig:window_office}.}.  Denoting the window loss by $l_w$, the signal from a BS in room $(i,j,k),k\neq 0$ to the typical window room $(0,0,0)$ is $hl_{|j|+|k|}l_w^2K_1^{|i|}$, where $h$ is the fading coefficient.

Now consider a signal path from a BS on the floor of the typical window room. We consider two paths from room $(i,j,0)$ to $(0,0,0)$. The first one is the direct path which penetrates the walls between room $(i,j,0)$ which contains a BS and the typical window room. The second one is the indirect path which radiates outside of the building through a window room $(0,j,0)$ and then goes to $(0,0,0)$ through the outside. So, the path loss model from room $(i,j,0)$ to $(0,0,0)$ becomes $h_dK_1^{|i|}K_2^{|j|}+h_{i}l_{|j|}l_w^2K_1^{|i|}$ where $h_d$ and $h_i$ are channel fading coefficients. With this path loss model, the interference distribution and the success probability can be computed as in Section \ref{sec:typical_room} and \ref{sec:Success and Coverage Probability}.

Denote the interference measured in the typical window room $(0,0,0)$ by $\tilde{I}_{f}$ under Rayleigh fading. The Laplace transform of $\tilde{I}_f$ is{
\begin{eqnarray*}
\textstyle\mathcal{L}_{\tilde{I}_f}(s)&\textstyle=&\textstyle\prod_{i\in\mathbb{N}\cup\{0\}}\left({1+\frac{4\lambda_1}{\mu_1}\sum_{j\in\mathbb{Z}}\sum_{k\in\mathbb{Z}}(1-\frac{1}{1+sK_1^{|i|}K_2^{|j|}K_3^{|k|}}\frac{1}{1+sl_{|j|+|k|}l_w^2K_1^{|i|}})}\right)^{-1}\\
\textstyle&\textstyle\times&\textstyle\prod_{j\in\mathbb{Z}}\left({1+\frac{4\lambda_2}{\mu_2}\sum_{i\in\mathbb{N}\cup\{0\}}\sum_{k\in\mathbb{Z}}(1-\frac{1}{1+sK_1^{|i|}K_2^{|j|}K_3^{|k|}}\frac{1}{1+sl_{|j|+|k|}l_w^2K_1^{|i|}})}\right)^{-1}\\
\textstyle&\textstyle\times&\textstyle\prod_{k\in\mathbb{Z}}\left({1+\frac{4\lambda_3}{\mu_3}\sum_{i\in\mathbb{N}\cup\{0\}}\sum_{j\in\mathbb{Z}}(1-\frac{1}{1+sK_1^{|i|}K_2^{|j|}K_3^{|k|}}\frac{1}{1+sl_{|j|+|k|}l_w^2K_1^{|i|}})}\right)^{-1}.
\end{eqnarray*}}This is obtained by adjusting indices with the new path loss model in Proposition \ref{prop:laplace_dist_fading_lines}. Since $0^0=1$, the direct and indirect path losses can be combined as above.

\begin{figure}
    	\begin{center}
    		\epsfxsize=2.8in {\epsfbox{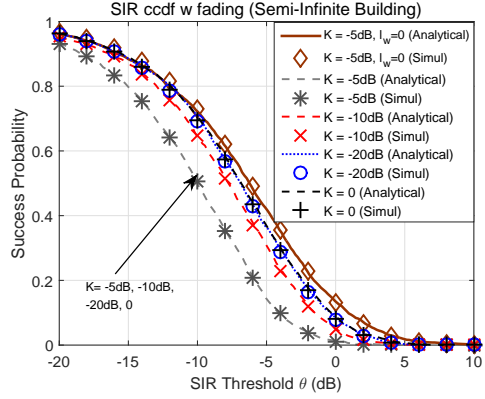}}
    		\caption{\small In-room link Success Probability (D2D transmission attempts from $(0,0,0)$ to itself) in the typical window room of semi-infinite Poisson building. $K_1=K_2=K$, $K_3=0$, $l_{|x|}=0.5^x$, $l_w=-3dB$ (except the reference curve without interference from outside the building), $r_i=\frac{\lambda_i}{\mu_i}=0.1$ for $i=1,2,3$, and $\sigma^2=0$. 
    			}
    		\label{fig:window_office_succ}
    	\end{center}
    \end{figure}

The success probability from room $(i,j,k)$ to room $(0,0,0)$ becomes
{
\begin{align*}
\textstyle\mathbb{P}[{\sf SINR}_s>\theta]=\mathcal{L}_{\tilde{I}_f}\left(\frac{\theta}{K_1^{|i|}K_2^{|j|}K_3^{|k|}}\right)\exp\left(-\frac{\theta\sigma^2}{K_1^{|i|}K_2^{|j|}K_3^{|k|}}\right).
\end{align*}}Fig. \ref{fig:window_office_succ} plots the in-room link success probability when $K_1,K_2=K$, $K_3=0$, $r_1,r_2,r_3=0.1$, $\sigma^2=0$, $l_{x}=0.5^x$, and $l_w = -3dB$. As $K$ decreases, the total interference from all transmitters decreases and the success probability increases. When $K=0$, all walls and floors block signals perfectly and interference comes through the outside of building only from transmitters in window offices. When we see the cases $K=0$, $K=-10dB$, and $K=-20dB$, the success probabilities are almost the same, which suggests that interference coming through the outside dominates the interference coming through the walls in the same floor. If we ignore the interference leakage from through OoB paths (\emph{i.e.,} $l_w=0$), the success probability for $K=-5dB$ is higher than $K=-20dB$ with interference from outside. This shows interference leakage through OoB paths is an important factor in the 3-D case.
  
\section{Conclusions}\label{sec:conclu}

We propose a Poisson grid based framework for analyzing urban indoor networks. The Poisson grid allows one to model general $n$-dimensional structures with randomly (but dependently) sized rooms, capturing the fact that more users are located in larger rooms. While the indoor propagation is dominated by shadowing and blockage, the framework facilitates the study of the correlated shadowing field and node distribution, which are one of the distinctive aspects of urban indoor geometry. The interference field associated to this environment is no more a shot noise field, because of common randomness by the shared static obstacles in the Poisson grid.

We obtain exact analytical expressions for the interference field and characterize the spectral efficiency of two basic communication scenarios in this context. We compare our correlated shadowing field and previous research (uncorrelated shadowing and free-space models) and observe very different moments and scaling laws. Thanks to its finite-size or semi-infinite size 3-D variants, the Poisson grid can be tailored for 1) computing the interference field given the building size and 2) analyzing interference leakage through the outside of the Poisson structure. 

This model opens a new way of analyzing urban indoor networks. Several variants can be derived from it to study more realistic architectural scenarios and wireless technologies.


\appendices 

\section{Proof of Proposition \ref{prop:moment_room_lines}}\label{appendix:moment_room_lines}
\lemma\label{lemma:geometric_sum} For all $K\in[0,1)$ and $k\in\mathbb{N}\cup\{0\}$, we have{
	\begin{eqnarray}
	\textstyle\sum_{i\in\mathbb{Z}}K^{|i|+|i-k|}= K^{|k|}\left(|k|+\frac{1+K^2}{1-K^2}\right)\mbox{, }~~~\sum_{(i,i')\in\mathbb{Z}^2}^{i\neq i'}K^{|i|+|i'-k|}=\left(\frac{1+K}{1-K}\right)^2-K^{|k|}\left(|k|+\frac{1+K^2}{1-K^2}\right).\nonumber
	\end{eqnarray}}Leveraging this lemma, we can obtain the joint moment of the interference between the typical room and room $(i_1,i_2,\ldots,i_n)$.

Let $N_{(i_1,i_2,\ldots,i_n)}$ be the number of the BSs in room $(i_1,i_2,\ldots,i_n)$. Since it is the sum of Poisson random variables, $\mathbb{E}[N_{(i_1,i_2,\ldots,i_n)}N_{(i_1',i_2',\ldots,i_n')}]=${
	\begin{align}
	\textstyle 2^{n-1}\left(\sum_{j=1}^n\frac{\lambda_j}{\mu_j}\right)\prod_{k=1}^n1_{i_k=i_k'}+2^{2n-2}\sum_{j=1}^n\frac{\lambda_j^2}{\mu_j^2}1_{i_j=i_j'}+2^{2n-2}\left(\sum_{j=1}^n\frac{\lambda_j}{\mu_j}\right)^2.\nonumber
	\end{align}}By leveraging Lemma \ref{lemma:geometric_sum}, we obtain $\mathbb{E}[I_{(0,0,\ldots,0)}I_{(i_1,i_2,\ldots,i_n)}]=$
	\begin{eqnarray}
	\textstyle&\textstyle&\textstyle\mathbb{E}[\sum_{(j_t)_{t=1}^n\in\mathbb{Z}^n}\prod_{m=1}^nK_m^{|j_m|}N_{(j_1,j_2,\ldots,j_n)}\sum_{(j_t')_{t'=1}^n\in\mathbb{Z}^n}\prod_{m=1}^nK_m^{|j_m'-i_m|}N_{(j_1',j_2',\ldots,j_n')}]\nonumber\\
	\textstyle&\textstyle=&\textstyle 	2^{n-1}\left(\sum_{j=1}^n\frac{\lambda_j}{\mu_j}\right)\left(\prod_{l=1}^nb_l(i_l)\right)+2^{2n-2}\left(\prod_{l=1}^na_l\right) \left(\left(\sum_{j=1}^n\frac{\lambda_j}{\mu_j}\right)^2 +\left(\sum_{j=1}^n\frac{\lambda_j^2b_j(i_j)}{\mu_j^2a_j}\right)\right)\nonumber.
	\end{eqnarray}

\section{Interference Laplace Transform Conditioned on $\delta = 0$ and $\Psi$}\label{appendix:nearest_conditioned}
In this section, we give the proof of the formula for coverage probability under the nearest room-distance BS association, conditioned on $\delta = 0$ and the given Poisson grid $\Psi$. First, we derive useful probabilities to obtain the interference Laplace transform.
	\begin{align*}
	\textstyle&\textstyle\mathbb{P}[\delta = 0|\Psi] = \sum_{n=1}^{\infty}\mathbb{P}[N=n|\delta = 0,\Psi]= 1-\exp(-\sum_{i=1}^{12}\lambda_i)\\
	\textstyle&\textstyle\mathbb{P}[N=1|\delta = 0,\Psi]  = (\sum_{i=1}^{12}\lambda_i)\times\exp(-\sum_{i=1}^{12}\lambda_i)/P_{tot}\\
	\textstyle&\textstyle\mathbb{P}[N=2|\delta = 0,\Psi]  = (\sum_{i=1}^{12}\frac{\lambda_i^2}{2!}+\sum_{(i,j)\in[1,12]}^{i< j}\lambda_i\lambda_j)\times\exp(-\sum_{i=1}^{12}\lambda_i)/P_{tot}
	\end{align*}Combining these results, the interference Laplace transform becomes 
	\begin{align*}
	\textstyle \mathbb{E}[e^{-sI}|\delta = 0,\Psi]&\textstyle=\sum_{n=1}^{\infty}\mathbb{E}[e^{-sI}|N=n,\delta = 0,\Psi] \mathbb{P}[N=n|\delta = 0,\Psi] \\
	\textstyle&	\textstyle=\sum_{n=1}^{\infty}\exp(-s\sum_{j=1}^{n-1}h_j)\mathbb{P}[N=n|\delta = 0,\Psi]\\
	\textstyle&\textstyle\times \prod_{i\in\mathbb{N},j,k\in\mathbb{Z}, i+|j|+|k|\geq 2}\exp(-4\lambda_1(x_i+x_{-i})(1-shK^{|i|+j+|k|-1}))\\
	\textstyle&\textstyle\times\prod_{j\in\mathbb{N},i,k\in\mathbb{Z}, |i|+j+|k|\geq 2}\exp(-4\lambda_2(y_j+y_{-j})(1-shK^{|i|+j+|k|-1}))\\
	\textstyle&\textstyle\times\prod_{k\in\mathbb{N},i,j\in\mathbb{Z}, |i|+|j|+k\geq 2}\exp(-4\lambda_3(z_k+z_{-k})(1-shK^{|i|+j+|k|-1})),
	\end{align*}where $x_i, y_j,z_k$ are the dimension of room $(i,j,k)$. The condition $i+|j|+|k|\geq 2$ (or $|i|+j+|k|\geq 2, |i|+|j|+k\geq 2$) implies considering the interference from the BSs out of the nearest room ($\delta = 0$). By deconditioning w.r.t. the channel coefficients $h\sim\exp(1)$, we obtain
	\begin{align*}
	\textstyle\mathbb{E}[e^{-sI}|\delta = 0,\Psi] 
	&\textstyle=\sum_{n=1}^{\infty}\left(\frac{1}{1+s}\right)^{n-1}\mathbb{P}[N=n|\delta = 0,\Psi]\\
	\textstyle&\textstyle\times \prod_{i\in\mathbb{N},j,k\in\mathbb{Z}, i+|j|+|k|\geq 2}\exp(-2\lambda_h(x_i+x_{-i})(1-\frac{1}{1+sK^{|i|+j+|k|-1}}))\\
	\textstyle&\textstyle\times\prod_{j\in\mathbb{N},i,k\in\mathbb{Z}, |i|+j+|k|\geq 2}\exp(-2\lambda_h(y_j+y_{-j})(1-\frac{1}{1+sK^{|i|+j+|k|-1}}))\\
	\textstyle&\textstyle\times\prod_{k\in\mathbb{N},i,j\in\mathbb{Z}, |i|+|j|+k\geq 2}\exp(-4\lambda_s(z_k+z_{-k})(1-\frac{1}{1+sK^{|i|+j+|k|-1}})).
	\end{align*}
	
\section{Proof of Proposition \ref{prop:finite_building}}\label{appendix: finite_building}
\lemma\label{lemma:finite_size} Let $t_1,t_2,\cdots,t_{n-1}\in\mathbb{R}$ be $n-1$ i.i.d. random variables uniformly distributed on $[0,d]$, where $d\in\mathbb{R}^+$. These $n-1$ points divide $[0,d]$ into $n$ intervals of length $y_1,y_2\cdots,y_n$, where $y_1$ is the length of left-most interval and $y_n$ is the right-most one. The Laplace transform of $(y_1,y_2,\ldots,y_n)$ is{	
\begin{align*}
\textstyle\mathcal{L}_{y_1,y_2,\cdots,y_n}(s_1,s_2,\ldots,s_n)=\frac{(n-1)!}{d^{n-1}}\frac{1}{2\pi j}\int_{c-j\infty}^{c+j\infty}e^{zd}\prod_{i=1}^n\frac{1}{z+s_i}dz,
\end{align*}}where $c>\mathcal{R}\{s_i\},\forall i\in[n]$.\begin{IEEEproof}
	The Laplace transform is $\mathcal{L}_{y_1,y_2,\ldots,y_n}(s_1,s_2,\ldots,s_n)=\mathbb{E}\prod_{i=1}^n e^{-s_iy_i}$. Since the random variables $\{t_i\}$ are i.i.d. with pdf $f_{t_i}(x)=\mathds{1}_{[0,d]}(x)/d$, the $n!$ possible orders of the $n-1$ random variable happen with equal probability and the Laplace transform can be written as {
\begin{align*}
\textstyle\mathcal{L}_{y_1,y_2,\ldots,y_n}(s_1,s_2,\ldots,s_n)=\frac{(n-1)!}{d^{n-1}}\int_{0}^d\int_{0}^{x_{n-2}}\int_{0}^{x_{n-3}}\cdots\int_{0}^{x_2}\prod_{i=1}^n e^{-s_i(x_i-x_{i-1})}dx_1dx_2\ldots dx_{n-1},
\end{align*}}where $x_0=0$ and $x_n=d$. Let $g_i(x)=e^{-s_ix}$. Then,{
\begin{eqnarray*}
\textstyle\mathcal{L}_{y_1,y_2,\ldots,y_n}(s_1,s_2,\ldots,s_n)=\frac{(n-1)!}{d^{n-1}}g_1*g_2*\cdots g_n(d),
\end{eqnarray*}}where $*$ denotes convolution. Since $\mathcal{L}_{g_i}(z)=\frac{1}{z+s_i}$, we have{
\begin{align*}
\textstyle g_1*g_2*\cdots *g_n(x)=\frac{1}{2\pi j}\int_{c-j\infty}^{c+j\infty}e^{zd}\prod_{i=1}^n\frac{1}{z+s_i}dz,
\end{align*}}where we made use of Mellin's inversion formula for the inverse Laplace transform.
\end{IEEEproof}
Due to the symmetricity of our network model, we only consider interference power coming from the BSs on the line segments which are parallel and negative direction to the $v_1$-axis and then extend the result without loss of generality. Let $\{\bar{x_i}\}$ be the length of line segments divided by $v_1$-orthogonal walls. Given $d_{v_{11}}$ and $n_{v_{11}}$ which is the number of $v_1$-orthogonal walls including the outermost walls, the interference distribution from the transmitters on line segments which are parallel to the $v_1$-axis and located on the $+v_1$ direction (which is denoted by $\tilde{I}_{f_{11}}$) becomes {
\begin{eqnarray*}
\textstyle	\mathbb{E}[e^{-\tilde{I}_{f_{11}}}|d_{v_{11}},n_{v_{11}},\{\bar{x}_i\}]=\prod_{i=1}^{n_{v_{11}}}\exp\left(-4\lambda_1\bar{x}_i\sum_{j=-n_{v_{21}}+1}^{n_{v_{22}}-1}\sum_{k=-n_{v_{31}}+1}^{n_{v_{32}}-1}(1-\frac{1}{1+sK_1^{|i|-1}K_2^{|j|}K_3^{|k|}})    \right),
\end{eqnarray*}	}and by deconditioning w.r.t. $\{\bar{x}_i\}$, 
\begin{align*}
&\textstyle\mathbb{E}[e^{-\tilde{I}_{f_{11}}}|d_{v_{11}},n_{v_{11}}]=\\
&\textstyle\frac{(n_{v_{11}}-1)!}{d_{v_{11}}^{n_{v_{11}}-1}}\frac{1}{2\pi j}\int_{c-j\infty}^{c+j\infty}e^{zd_{v_{11}}}\prod_{i=1}^{n_{v_{11}}}   \frac{1}{z+4\lambda_1\sum_{j=-n_{v_{21}}+1}^{n_{v_{22}}-1}\sum_{k=-n_{v_{31}}+1}^{n_{v_{32}}-1}(1-\frac{1}{1+sK_1^{|i|-1}K_2^{|j|}K_3^{|k|}})}      dz \mbox{,}
\end{align*}where $n_{v_{21}}, n_{v_{22}}, n_{v_{31}},n_{v_{32}}$ are the numbers of walls along the $-v_2,v_2,-v_3$ and $v_3$ directions respectively. Since the probability mass function of $n_{v_{11}}$ is $f_{n_{v_{11}}}(k)=\frac{{(\mu_1d_{v_{11}})}^{k-1}e^{-\mu_1d_{v_{11}}}}{(k-1)!}$,  by deconditioning w.r.t. $n_{v_{11}}$ and considering interference from all transmitters, we obtain the result of Proposition \ref{prop:finite_building} when we consider all BSs in a finite size Poisson building.

\section*{Acknowledgement}
This research was funded by the NSF CCF CIF Medium grant $(\# 1514275)$ "Fundamentals of Urban Millimeter Wave Networks" to The University of Texas at Austin and by a Math+X award from the Simons Foundation $(\# 197982)$ to The University of Texas at Austin. It was also supported by Samsung through a grant to the Wireless Network and Communication Group of The University of Texas at Austin.

\bibliographystyle{Bibs/ieeetran}
\bibliography{Bibs/IEEEabrv,Bibs/referenceBibs,Bibs/xinchen}

\end{document}